\documentclass[aps,prc,preprint,superscriptaddress]{revtex4-1}
\usepackage{graphicx,epsfig,epstopdf} 
\begin{document}

\title{Pauli resonance states in light nuclei: how they appear and how they can
be eliminated}
\author{N. Kalzhigitov}
\email{knurto1@gmail.com}
\affiliation{Al-Farabi Kazakh National University, Almaty, Kazakhstan}
 
\author{V. S. Vasilevsky}
\email{vsvasilevsky@gmail.com}
\affiliation{Bogolyubov Institute for Theoretical Physics,\\
 Kyiv, 03143, Ukraine
} 
%\author{N. Kalzhigitov$^{1}$, V. S. Vasilevsky$^{2}$
%\and $^{1}$Al-Farabi Kazakh National University, Almaty, Kazakhstan
%\and $^{2}$Bogolyubov Institute for Theoretical Physics, Kyiv, Ukraine}
%\maketitle

%\keywords{Resonating group method, light nuclei, Pauli principle,  Pauli resonance states, Pauli allowed states, norm kernel}%Use showkeys class option if keyword

\date{\today}

\begin{abstract}
A systematic analysis of parameters and properties of the Pauli resonance states
 is performed for light nuclei $^{6}$Li, $^{7}$Li, $^{8}$Be, $^{9}$Be and
$^{10}$B, which are treated as two-cluster systems. The Pauli resonance states
are redundant solutions of the resonating group method. They appear, when one try to
use a more advanced description of the internal structure of interacting
clusters. Our calculations are performed in the standard and advanced versions
of the resonating group method. The standard version employs wave functions of
the many-particle shell model to describe the internal motion of nucleons within each
cluster. The advanced version is based on a three-cluster resonating group
method. As in the standard version, the internal wave functions of three
clusters are approximated by wave functions of the many-particle shell
model. However, in the advanced version, a pair of clusters can form a bound
state, and then the third cluster is considered to interact with such two clusters, being 
in such the bound state. It is
found that the Pauli resonance states in nuclei under consideration are observed 
at energies between 11 and 46 MeV, and their widths vary from 8 keV to 6.7 MeV.
The analysis of the wave functions of Pauli resonance states and matrix elements of the
norm kernel allowed us to formulate an effective method for eliminating Pauli
resonance states. It is demonstrated that this method effectively eliminates
all determined the Pauli resonance states.

\end{abstract}
\maketitle

%\keywords{}


\section{Introduction}

We are going to study properties of the so-called Pauli resonance states which
have been numerously observed in Refs. \cite{1974PhLB...49..308C,
1981ZPhyA.303..209S, 1975PThPh..54.1707K,
1975PThPh..54..747K, 1982NuPhA.377...84F,
1982NuPhA.380...87K, 1982PhRvC..26.2410S,
1983PhRvC..27.2457F, 1983NuPhA.394..387W,
1985NuPhA.437..367W, 1988PhRvC..38.2013K,
1992NuPhA.548...39K} and many others. These resonance states appear
within the resonating group method (RGM), when one tries to use a more realistic
description of interacting nuclei (clusters). They have been considered as redundant solutions of the equations of the resonating group method. As the Pauli resonance states
appear not in all realizations (versions) of the resonating group method, we
start with a short classification of main versions of the RGM, which are
relevant to the subject of the present paper. The main difference of these
methods is in the form of the wave function, which is used to
approximate the cluster structure of a compound nucleus. The standard version of
the RGM suggests the following form of the wave function of a two-cluster system
of $A$ nucleons for the partition $A=A_{1}+A_{2}$:
\begin{equation}
\Psi\left(  A\right)  =\widehat{\mathcal{A}}\left\{  \Phi_{1}\left(
A_{1},b\right)  \Phi_{2}\left(  A_{2},b\right)  \psi\left(  \mathbf{x}\right)
\right\}  , \label{eq:I01}%
\end{equation}
where $\mathbf{x}$ is a distance between centers of mass of clusters, 
$\psi\left(  \mathbf{x}\right)$ is the wave function of the relative motion of clusters,
$\Phi_{1}\left(  A_{1},b\right)  $ and $\Phi_{2}\left(  A_{2},b\right)  $ are the
wave functions of the many-particle shell model describing the motion of nucleons
within the first and second clusters, respectively. They are antisymmetric and
translationally invariant. The oscillator length $b$ determines the effective sizes of clusters. An important component of Eq. (\ref{eq:I01}) is the antisymmetrization operator $\widehat{\mathcal{A}}$ which makes the antisymmetric wave function of a compound system.
For the sake of brevity, we omit all quantum numbers. They will be explicitly
indicated in Sec. \ref{Sec:Method}.

In the second version, which we call the improved one, the wave function is chosen in the
form%
\begin{equation}
\Psi\left(  A\right)  =\widehat{\mathcal{A}}\left\{  \Phi_{1}\left(
A_{1},b_{1}\right)  \Phi_{2}\left(  A_{2},b_{2}\right)  \psi\left(
\mathbf{x}\right)  \right\}  , \label{eq:I02}%
\end{equation}
where different oscillator lengths $b_{1}$ and $b_{2}$ are used to improve the
description of the internal structure of each cluster. This version is
suitable for clusters with large difference of masses, i.e., for example, when
$A_{1}\gg A_{2}$.

The third version is called the advanced version of the RGM and related to
the advanced description of the internal structure of one cluster
\begin{equation}
\Psi\left(  A\right)  =\widehat{\mathcal{A}}\left\{  \Phi_{1}\left(
A_{1},b_{1}\right)  \Psi_{2}\left(  A_{2},b\right)  \psi\left(  \mathbf{x}%
\right)  \right\}  , \label{eq:I03}%
\end{equation}
or two clusters%
\begin{equation}
\Psi\left(  A\right)  =\widehat{\mathcal{A}}\left\{  \Psi_{1}\left(
A_{1},b\right)  \Psi_{2}\left(  A_{2},b\right)  \psi\left(  \mathbf{x}\right)
\right\}  . \label{eq:I03A}%
\end{equation}
Contrary to the wave function $\Phi_{\alpha}\left(  A_{\alpha},b\right)  $
($\alpha$=1,2), the wave function $\Psi_{\alpha}\left(  A_{\alpha},b\right)  $ is
a solution of the two-cluster Schr\"{o}dinger equation with clusterization
$A_{\alpha}=A_{\alpha1}+A_{\alpha2}$ and is presented in the form similar to
(\ref{eq:I01}). This version of the RGM suggests a more correct description of the
compound system and is appropriate, when one or both clusters $A_{1}$ and
$A_{2}$ have an evident two-cluster structure, or, in other
words, they have weakly bound state(s) and thus can be easily split into two
fragments. Many light nuclei such as $d$, $^{6}$Li, $^{7}$Li, and $^{7}$Be have
such properties as their separation energies are less than 3 MeV.

The Pauli resonance states have not seen in the standard version of the RGM. Only the shape
resonance states were detected within this version in the single-channel
approximation. As is well known, the shape resonances are created by the
centrifugal and/or Coulomb barriers. Thus, they lie relatively close to the
threshold of the corresponding channel. 

The Pauli resonance states have been
detected in the improved and advanced versions of the RGM.
The most spectacular demonstration of the Pauli resonance states was presented
in Refs. \cite{1983NuPhA.394..387W, 1985NuPhA.437..367W}, where the elastic
scattering of alpha particles on $^{16}$O has been calculated within the standard
and improved versions. A set of narrow and wide resonance states was emerged,
when different oscillator lengths (frequencies) were used for the wave functions
describing the internal structure of $^{16}$O and $^{4}$He nuclei. They spread
over a wide energy interval from small to relatively high energies above the
$^{16}$O+$^{4}$He threshold. For example, in Ref. \cite{1983NuPhA.394..387W}, 
in the $0^{+}$ state, the improved version generates four resonance states
with the energy less than 30 MeV, while the standard version creates 
no resonance states. These results have been obtained with the delta-shape
nucleon-nucleon potential. In Ref. \cite{1985NuPhA.437..367W}, the same models
were used for the resonance structure in the $^{16}$O+$^{4}$He scattering, but
with a semirealistic nucleon-nucleon potentials. It was found that, in this
case, the Pauli resonances are shifted to the high-energy region ($E>$30 MeV).

In light nuclei, within the advanced version of
the RGM \cite{1974PhLB...49..308C, 1981ZPhyA.303..209S,
 1975PThPh..54.1707K, 1975PThPh..54..747K,
 1982NuPhA.380...87K, 1982PhRvC..26.2410S,
 1983PhRvC..27.2457F,1988PhRvC..38.2013K}, the Pauli resonance
states have been observed in a relatively high energy region $E>$%
15 MeV. It was also noticed in Ref. \cite{1981ZPhyA.303..209S} that the Pauli
resonance states manifest themselves in the states with small values of the
total orbital momentum $L$.

Despite that the different authors have used different names for such type of
resonances, such as "positive energy bound states" \cite{1974PhLB...49..308C},
"redundant" \cite{1982NuPhA.377...84F} or "spurious states"
\cite{1982PhRvC..26.2410S}, it was widely recognized that the correctly
treated Pauli principle is the origin of those states.

Why are the Pauli resonances considered as spurious states? There are two
reasons for treating such resonance states. First, there are no physical
justifications for the appearance of Pauli resonances. Second, such resonances
are not observed experimentally.

To clarify the first reason, let us recall the main types of resonance states are
observed in many-particle and particular in nuclear systems.
 The first type is shape resonance states
which are created by centrifugal or/and Coulomb barriers. The second type is
represented by the Feshbach resonance states. These resonance states appear
due to a weak coupling between open and closed channels. There are two
necessary conditions for creating the Feshbach resonances. A compound system
should have at least two channels with different threshold energies, and
there should be at least one bound state in the channel with the larger threshold
energy provided that this channel is considered separately from the channel
with the lowest threshold energy.

The phenomenon which is called the Pauli resonance state cannot be explained
by two main factors creating resonance states and, thus, cannot be
attributed to the first or second type of resonances. It cannot be the
Feshbach resonance as such resonance states observed in single-channel cases. 
It is impossible to relate the Pauli resonances to
centrifugal or Coulomb barrier, since they appear in states with zero or very
small angular momenta, or they require a very huge barrier.

To understand the phenomenon of Pauli resonance states, it is worth to
recall some properties of the antisymmetrization operator $\widehat
{\mathcal{A}}$. First, this operator been applied to a many-particle function
makes this function totally antisymmetric with respect to a permutation of any
pair of particles or annihilates it. The latter means that such wave function
cannot be antisymmetric. It is said that the Pauli principle prohibits such
function or makes it forbidden. Usually, such type of functions describes
many-particle systems, when more than four nucleons occupy the same
single-particle orbital. Second, the antisymmetrization operator may
significantly affect the normalization properties of many-particle functions.
If the operator  $\widehat{\mathcal{A}}$ is applied to a wave function, which
is normalized to unity, then the overlap of the resulting  antisymmetric wave
function can be smaller than unity, larger than unity or even very small.

Both properties of the antisymmetrization operator $\widehat{\mathcal{A}}$
hava a great impact on the structure of equations for many-cluster systems, on
the explicit form and the interpretation of obtained solutions. These properties are
discussed in more details in the following  sections of the paper.

The Pauli resonance states have been considered as redundant solutions of the
RGM equations. Thus, one needs to use an algorithm to eliminate
these states. They distort real physical quantities such as the phase shifts,
cross-sections of various processes, and so on. To the best of our knowledge, there is
only one algorithm for eliminating the Pauli resonance states. It was formulated in
Ref. \cite{1992NuPhA.548...39K}  and applied to the $^4$He+$^{16}$O system.
We refer to this method as the REV method, which removes the eigenvalues of 
the norm kernel that cause the Pauli resonances. It was suggested in Ref.
\cite{1992NuPhA.548...39K} to omit the so-called almost forbidden Pauli states.
The criterion how to distinguish such states from allowed ones was formulated.
This algorithm has eliminated all Pauli resonance states from the
elastic scattering of alpha particles on $^{16}$O.

In the present paper, we are going to examine the continuous-spectrum states of a
set of light nuclei such as $^{6}$Li, $^{7}$Li, $^{7}$Be, $^{8}$Be, $^{9}$Be
and $^{10}$B. All these nuclei are considered as a three-cluster configuration
and treated within a three-cluster model formulated in Ref.
\cite{2009NuPhA.824...37V}. The three-cluster configuration is then reduced to
three (if all three clusters are different) or two (if two of three clusters
are identical) binary channels. When such reduction causes a pair of clusters to form a bound state, this state is described by the two-cluster approximation in our method.
%With such reduction, one pair of clusters
%forms a bound state, which within our method is described in a two-cluster approximation.

To study the Pauli resonance states, we first analyze the overlap matrix 
and its eigenvalues for several light nuclei. Based on this analysis, we
suggest an alternative method for eliminating the Pauli resonance states. 
We call this new method of Removing of the Oscillator Functions the ROF method.
%This
%new method we call as the ROF method, it means Removing of Oscillator
%Functions.
 We will demonstrate that both methods of REV and ROF give close results and
completely eliminate all Pauli resonance states.

The structure of the present paper is as follows. In Section
\ref{Sec:Method}, we give a brief introduction of the methods applied to study
properties of the Pauli resonances in light nuclei. In Section
\ref{Sec:Results}, the choice of input parameters and the details of calculations
are discussed. The manifestations of the Pauli resonance states in various
two-cluster systems are demonstrated in Section \ref{Sec:Manifest}. The
analysis of the parameters of Pauli resonance states and their wave functions
is carried out in this section. Then in the Section \ref{Sec:Overlap}, we
analyze the matrix elements and the eigenvalues of a norm kernel.
 In Section \ref{Sec:REV}, we briefly explain the main idea of eliminating
the Pauli resonance states suggested in Ref. \cite{1992NuPhA.548...39K}. Here,
we also demonstrate its efficiency. In Section \ref{Sec:ROF}, we formulate an
alternative method for eliminating the Pauli resonance states and demonstrate
how it works in a two-cluster systems under consideration. Concluding remarks
are presented in Section \ref{Sec:Conclus}.

\section{Method \label{Sec:Method}}

In this paper we will use two types of two-cluster functions and, thus, two
realizations of the RGM. The first type of functions represents the standard
form of the resonating group method, and the second type realizes the advanced
form of the RGM. The wave function of the first type for a partition $A=A_{1}+A_{2}$
reads
\begin{equation}
\Psi_{E,J}\left(  A\right)  =\widehat{\mathcal{A}}\left\{  \left\{  \left[
\Phi_{1}\left(  A_{1},L_{1},S_{1},b\right)  \Phi_{2}\left(  A_{2},L_{2}%
,S_{2},b\right)  \right]  _{S}\psi_{E,l,L,J}\left(  x\right)  Y_{l}\left(
\widehat{\mathbf{x}}\right)  \right\}  _{L}\right\}  _{J}, \label{eq:M01}%
\end{equation}
and the wave functions of the second type for the partition $A=A_{1}+A_{2}%
=A_{1}+\left(  A_{21}+A_{22}\right)  $ are%
\begin{equation}
\Psi_{E,J}\left(  A\right)  =\widehat{\mathcal{A}}\left\{  \left\{  \left[
\Phi_{1}\left(  A_{1},L_{1},S_{1},b\right)  \Psi_{2}\left(  A_{2},L_{2}%
,S_{2},b\right)  \right]  _{S}\psi_{E,l,J}\left(  x\right)  Y_{l}\left(
\widehat{\mathbf{x}}\right)  \right\}  _{L}\right\}  _{J}, \label{eq:M02}%
\end{equation}
where $\Psi_{2}\left(  A_{2},S_{2},L_{2},b\right)  $ is the wave function of a
bound state of the two-cluster subsystem with the partition $\left(  A_{21}%
+A_{22}\right)  $%
\begin{equation}
\Psi_{2}\left(  A_{2},L_{2},S_{2},b\right)  =\widehat{\mathcal{A}}\left\{
\left[  \Phi_{1}\left(  A_{21},S_{21},b\right)  \Phi_{2}\left(  A_{22}%
,S_{22},b\right)  \right]  _{S_{2}}g_{\mathcal{E},\lambda,J}\left(  y\right)
Y_{\lambda}\left(  \widehat{\mathbf{y}}\right)  \right\}  _{J}. \label{eq:M03}%
\end{equation}
Recall that we use the capital letter $\Phi$ to denote the wave
functions that are not solutions of the corresponding Schr\"{o}dinger equation,
they are the wave functions of the many-particle shell model. These functions can
be constructed as Slater determinants from single-particle oscillator
orbitals. The capital and small letters $\Psi$ and $\psi$ represent solutions
of the many-particle Schr\"{o}dinger equation or corresponding
integro-differential Wheeler equation \cite{1937PhRv...52.1083W,
 1937PhRv...52.1107W}.

Here, we use the $LS$ coupling scheme, when the total spin $S$ is a
vector sum of spins of clusters, and the total orbital momentum $L$ is a vector
sum of the orbital momenta of both clusters $L_{1}$ and $L_{2}$ and the orbital
moment of the relative motion of clusters $l$.

In the present paper, we consider the special case of the advanced version of the
RGM. The usage of the special case is justified by employing  the three-cluster
model for investigating the cluster-cluster scattering and the structure of  a 
compound nucleus. In this special case, only one of two functions 
$\Psi_{1}$ and $\Psi_{2}$ of the internal motion of nucleons
is a solution of the two-cluster Schr\"{o}dinger equation, and another function is
the many-body shell-model wave function. A four-cluster model will allow one
to consider the general case with two wave functions $\Psi_{1}$ and $\Psi_{2}$
to be solutions of the two-cluster Schr\"{o}dinger equations.

To realize the advanced model, we employ the three-cluster model which was proposed
in Refs. \cite{2009NuPhA.824...37V, 2009PAN....72.1450N}. Within this
model, a three-cluster configuration is transformed into a set of binary
channels, i.e., in several pairs of interacting nuclei, and one of the
interacting nuclei is considered as a two-cluster system. In Refs.
\cite{2009NuPhA.824...37V, 2009PAN....72.1450N}, the model has been
applied to study nuclei $^{7}$Be and $^{7}$Li with three-cluster
configurations $^4$He+$d+n$ and $^4$He+$d+p$, respectively. The  structure of
the $^{10}$B nucleus has been investigated in Ref. \cite{2014UkrJPh..59.1065N}
by employing the three-cluster configuration $^4$He+$^4$He+$d$. Recently, the
model which involves two three-cluster configurations $^4$He+$p+n$ and $^3$H+$d+p$
was used in Ref. \cite{KALZIGITOV2021APP} to study the resonance states of
$^{6}$Li in a wide energy range.

The model involves the Gaussian basis functions to determine bound-state wave functions
of two-cluster subsystems and the oscillator basis functions to describe the scattering of the
third cluster on a bound state of the two-cluster subsystem. The abbreviation
AMGOB is used to distinguish this model. In the AMGOB, two-cluster
(\ref{eq:M03}) and three-cluster (\ref{eq:M02}) wave functions are represented
as%
\begin{eqnarray}
& & \Psi_{2}\left(  A_{2},S_{2},L_{2},b\right)  \nonumber \\
& & =\sum_{\nu=1}^{N_{G}}D_{\nu
}^{E,L_{2},J}\widehat{\mathcal{A}}\left\{  \left[  \Phi_{1}\left(  A_{21}%
,S_{21},b\right)  \Phi_{2}\left(  A_{22},S_{22},b\right)  \right]  _{S_{2}}%
G_{L_{2}}\left(  x,b_{\nu}\right)  Y_{L_{2}}\left(  \widehat{\mathbf{x}}\right)
\right\}  _{J},\label{eq:M04}\\
&&  \Psi_{E,J}\left(  A\right)  =\sum_{n=0}^{N_{O}}C_{nL}^{E,J}\widehat
{\mathcal{A}}\left\{  \left[  \Phi_{1}\left(  A_{1},S_{1},b\right)  \Psi
_{2}\left(  A_{2},S_{2},L_{2},b\right)  \right]  _{S,L_{2}}\Phi_{n,L}\left(
y,b\right)  Y_{L}\left(  \widehat{\mathbf{y}}\right)  \right\}  _{J},
\label{eq:M05}%
\end{eqnarray}
where
\begin{equation}
G_{L}\left(  x,b_{\nu}\right)  =\frac{1}{b_{\nu}^{3/2}}\sqrt{\frac{2}%
{\Gamma\left(  L+3/2\right)  }}\rho^{L}\exp\left\{  -\frac{1}{2}\rho
^{2}\right\}  ,\qquad\left(  \rho=\frac{x}{b_{\nu}}\right)  , \label{eq:M031}%
\end{equation}
is the Gaussian function, and
\begin{eqnarray}
\Phi_{n,L}\left(  y,b\right)   &  =& \left(  -1\right)  ^{n}\mathcal{N}%
_{nL}~b^{-3/2}\rho^{L}e^{-\frac{1}{2}\rho^{2}}L_{n}^{L+1/2}\left(  \rho
^{2}\right)  , \label{eq:M032}\\
\rho &  =&\frac{y}{b},\quad\mathcal{N}_{nL}=\sqrt{\frac{2\Gamma\left(
n+1\right)  }{\Gamma\left(  n+L+3/2\right)  }}, \nonumber
\end{eqnarray}
is the oscillator function. In Eqs. (\ref{eq:M031}) and (\ref{eq:M032}),
$b_{\nu}$ and $b$\ denote oscillator lengths. The motivation to use these
functions can be found in Ref. \cite{2009NuPhA.824...37V}. The expansion
coefficients $D_{\nu}^{E,L_{2},J}$ and $C_{nL}^{E,J}$ are solutions of a set of
linear equations originated from the corresponding Schr\"{o}dinger equations. This
is a system of equations for the expansion coefficients $D_{\nu}^{E,L_{2},J}$\
\begin{equation}
\sum_{\widetilde{\nu}=0}\left[  _{G}\left\langle \nu,L_{2}\left\vert \widehat
{H}^{\left(  2\right)  }\right\vert \widetilde{\nu},L_{2}\right\rangle_{G}
-E\ _{G}\left\langle \nu,L_{2}|\widetilde{\nu},L_{2}\right\rangle_{G} \right]  D_{\widetilde
{\nu}}^{E,L_{2},J}=0, \label{eq:M09}%
\end{equation}
and we have a system of equations for the expansion coefficients $C_{nL}%
^{E,J}$:%
\begin{equation}
\sum_{\widetilde{n}=0}\left[  \left\langle n,L\left\vert \widehat
{H}\right\vert \widetilde{n},L\right\rangle -E\left\langle n,L|\widetilde
{n},L\right\rangle \right]  C_{\widetilde{n}L}^{E,J}=0. \label{eq:M10}%
\end{equation}
The system of equations (\ref{eq:M09}) involves matrix elements of the two-cluster
Hamiltonian 
\begin{equation}
_{G} \left\langle \nu,L_{2}\left\vert \widehat{H}^{\left(  2\right)
}\right\vert \widetilde{\nu},L_{2}\right\rangle_{G}  \nonumber
\end{equation} 
and the unit operator (norm kernel)
$_{G} \left\langle \nu,L_{2}|\widetilde{\nu},L_{2}\right\rangle_{G} $ between cluster Gaussian
functions%
\begin{equation}
\left\vert \nu,L_{2}\right\rangle_{G} =\widehat{\mathcal{A}}\left\{   \left[  \Phi_{1}\left(  A_{21}%
,S_{21},b\right)  \Phi_{2}\left(  A_{22},S_{22},b\right)
\right]  _{S_{2}} G_{L_{2}}\left(  x,b_{\nu}\right)  Y_{L_{2}}\left(  \widehat
{\mathbf{y}}\right)  \right\}  _{J}, \label{eq:M110}%
\end{equation}
while the system of equations (\ref{eq:M10}) involves matrix elements of the
three-cluster Hamiltonian $\left\langle n,L\left\vert \widehat{H}\right\vert
\widetilde{n},L\right\rangle $ and the unit operator $\left\langle n,L|\widetilde
{n},L\right\rangle $ between cluster oscillator functions%
\begin{equation}
\left\vert n,L\right\rangle =\widehat{\mathcal{A}}\left\{  \left[  \Phi
_{1}\left(  A_{1},S_{1},b\right)  \Psi_{2}\left(  A_{2},S_{2},L_{2},b\right)
\right]  _{S,L_{2}}\Phi_{n,L}\left(  y,b\right)  Y_{L}\left(  \widehat
{\mathbf{y}}\right)  \right\}  _{J}. \label{eq:M11}%
\end{equation}
 We will also use another basis of cluster oscillator functions%
\begin{equation}
\left\vert n,L\right\rangle _{0}=\widehat{\mathcal{A}}\left\{  \left[
\Phi_{1}\left(  A_{1},S_{1},b\right)  \Phi_{2}\left(  A_{2},S_{2}%
,L_{2},b\right)  \right]  _{S,L_{2}}\Phi_{n,L}\left(  y,b\right)  Y_{L}\left(
\widehat{\mathbf{y}}\right)  \right\}  _{J} \label{eq:M13}%
\end{equation}
to expand the wave functions of two-cluster systems in the standard version of the
RGM (\ref{eq:M01}). It is obvious that the wave functions $\left\vert
n,L\right\rangle _{0}$ are a partial case of wave functions $\left\vert
n,L\right\rangle $, when the second cluster has the most compact shape.

As was pointed out in Introduction, the Pauli principle plays the paramount role
in nuclear systems. The Pauli principle is realized through the
antisymmetrization operator  $\widehat{\mathcal{A}}$. Two main properties of
the operator were mentioned in Introduction. Now, we consider
the second property - the action of the antisymmetrization operator on the
normalization properties of wave functions. Consider, for example, the wave
functions (\ref{eq:M11}) and (\ref{eq:M13}). Each function in (\ref{eq:M11})
and (\ref{eq:M13}) to the right of the antisymmetrization operator is
normalized to unity. However, the antisymmetric functions in the general case are not
normalized to unity, as we shall see below. The overlap $\left\langle
n,L|n,L\right\rangle $ deviates from unity, when the quantum number $n$ is
small, or, in other words, when the distance between clusters is small. The
antisymmetrization operator makes the overlap $\left\langle
n,L|n,L\right\rangle $ larger than unity or smaller. In some cases, it makes
$\left\langle n,L|n,L\right\rangle $ very close to zero. Undoubtedly, these
properties of the antisymmetrization operator \ have to be taken into account,
when we are solving Eqs.  (\ref{eq:M10}).

The appearance of the matrix $\left\Vert \left\langle n,L|\widetilde
{n},L\right\rangle \right\Vert $ in Eq. (\ref{eq:M10}) indicates that the cluster
oscillator basis (\ref{eq:M11}) is not orthonormal, despite that all functions
to the right of the antisymmetrization operator in Eq. (\ref{eq:M11}) are
normalized to unity on the corresponding part of the coordinate space. This matrix
plays an important role in cluster models. It reflects effects of the Pauli
principle. If one neglects the total antisymmetrization by putting
$\widehat{\mathcal{A}}=1$, one obtains the unit matrix $\left\Vert \left\langle
n,L|\widetilde{n},L\right\rangle \right\Vert $. When the effects of the Pauli
principle are small, then the diagonal matrix elements are close to unity, and the
off-diagonal matrix elements tend to zero. Such behavior of the matrix elements
$\left\langle n,L|\widetilde{n},L\right\rangle $ is observed for large values
of $n$ and $\widetilde{n}$. This region of quantum numbers $n$ and
$\widetilde{n}$ corresponds to large distances between clusters and, thus, is
called the asymptotic region.

In the standard version of the RGM, the matrix $\left\Vert \left\langle
n,L|\widetilde{n},L\right\rangle \right\Vert $ is diagonal for two
$s$-clusters, as the orbital momenta of the first and second clusters
$L_{1}$=$L_{2}$=0. Within the advanced version of the RGM, as will be demonstrated
below, the matrix $\left\Vert \left\langle n,L|\widetilde{n},L\right\rangle
\right\Vert $ is not diagonal. However, the largest matrix elements are
situated on the main diagonal of the matrix.

It is worthwhile noticing that the wave functions $\left\{  C_{nL}^{E,J}\right\}
$ obtained by solving the system of equations (\ref{eq:M10}) are normalized by the
conditions
\begin{equation}
\sum_{n,n=0}C_{nL}^{E_{\alpha},J}\left\langle n,L|\widetilde{n},L\right\rangle
C_{\widetilde{n}L}^{E_{\alpha},J}=\delta_{\alpha\beta} \label{eq:M14A}%
\end{equation}
for states of the discrete spectrum and
\begin{equation}
\sum_{n,n=0}C_{nL}^{E,J}\left\langle n,L|\widetilde{n},L\right\rangle
C_{\widetilde{n}L}^{\widetilde{E},J}=\delta\left(  E-\widetilde{E}\right)
\label{eq:M14B}%
\end{equation}
for the continuous-spectrum states. An important consequence of Eqs.
(\ref{eq:M14A}) and (\ref{eq:M14B}) is that the value $\left\vert C_{nL}%
^{E,J}\right\vert ^{2}$ does not determine the contribution of the
oscillator functions $\left\vert n,L\right\rangle $ to the norm of a bound
state or continuous-spectrum state.

To solve Eq. (\ref{eq:M10}) for a finite number of basis functions
($n$=0, 1, 2, \ldots, $N_{O}-1$), one needs to analyze the $N_{O}\times N_{O}$
matrix $\left\Vert \left\langle n,L|\widetilde{n},L\right\rangle \right\Vert
$, whether this matrix contains redundant states which are called the Pauli
forbidden states. For this aim, the diagonalization procedure is usually
employed. It yields the eigenvalues $\Lambda_{\alpha}$ ($\alpha$=1, 2,
\ldots, $N_{O}$) and corresponding eigenfunctions $\left\Vert U_{n}^{\alpha
}\right\Vert $ of the matrix $\left\Vert \left\langle n,L|\widetilde
{n},L\right\rangle \right\Vert $. Eigenstates with $\Lambda_{\alpha}=0$ are
called the Pauli forbidden states and have to be removed from the space.
Eigenstates with small values of $\Lambda_{\alpha}$ are called the partially
or almost forbidden states. Usually, there are a large number of eigenstates
with $\Lambda_{\alpha}=1$. These states are not affected by the
antisymmetrization. Besides, the matrix $\left\Vert \left\langle
n,L|\widetilde{n},L\right\rangle \right\Vert $ can have eigenvalues with
$\Lambda_{\alpha}>1$. They are called the super allowed states.

Note that the construction of the Pauli allowed states is a key problem for
many-cluster systems. Many algorithms have been formulated (see, for example,
\cite{1977PThPS..62...90H, 1980PThPh..63..895F,
 1988PThPh..80..663K},) to construct and to select Pauli allowed states.

Actually, we have two different discrete
representations of the Schr\"{o}dinger equation. The first representation is
the oscillator basis representation and will be referred as the $n-$%
representation. The second representation is formed by eigenvalues of the norm
kernel matrix and will be referred as the $\alpha-$representation. We 
recall that both representations are related by the orthogonal matrix
$\left\Vert U_{n}^{\alpha}\right\Vert $.

In the $\alpha-$representation, the set of equations (\ref{eq:M10}) is
transformed to the form%
\begin{equation}
\sum_{\widetilde{\alpha}=1}^{N_{O}}\left[  \left\langle \alpha,L\left\vert
\widehat{H}\right\vert \widetilde{\alpha},L\right\rangle -E\Lambda_{\alpha
}\delta_{\alpha,\widetilde{\alpha}}\right]  C_{\widetilde{\alpha}L}^{E,J}=0,
\label{eq:M20}%
\end{equation}
where
\begin{equation}
\left\langle \alpha,L\left\vert \widehat{H}\right\vert \widetilde{\alpha
},L\right\rangle =\sum_{n,\widetilde{n}=0}^{N_{O}}U_{n}^{\alpha}\left\langle
n,L\left\vert \widehat{H}\right\vert \widetilde{n},L\right\rangle
U_{\widetilde{n}}^{\widetilde{\alpha}}. \label{eq:M21}%
\end{equation}
If the cluster system under consideration contains no the Pauli forbidden
states, then one may use the set of equations (\ref{eq:M10}) or (\ref{eq:M20}%
), both sets give the same spectrum, but different wave functions. One has to
use the set of equations (\ref{eq:M20}), when there are one or more the Pauli
forbidden states.

To study effects of the Pauli principle, we will analyze the overlap matrix
$\left\Vert \left\langle n,L|\widetilde{n},L\right\rangle \right\Vert $. We
will also analyze the eigenvalues and eigenfunctions of the matrix.

Few words about stages of solving the three-cluster problem. Before solving
the three-body or three-cluster equations, one has to solve the two-body or
two-cluster problem(s). The energies of bound states of a two-cluster subsystem
determine threshold energies which are of great importance for implementing
the proper boundary conditions for the three-cluster system. Thus, at the first stage,
we have to find the spectrum and wave functions of the two-cluster subsystem by
solving the generalized eigenvalue problem represented by Eq. (\ref{eq:M09}). 
To optimize calculations with the Gaussian wave functions, we
parametrize a set of widths $b_{\nu}$ with parameters $b_{0}$ and $q$ as%
\begin{equation}
b_{\nu}=b_{0}q^{\nu-1},\quad\nu=1,\ldots,N_{G}.
\label{eq:M22}
\end{equation}
The parameters $a_{0}$ and $q$ are used as variational parameters to minimize
the ground-state energy of the two-cluster subsystem. Such parametrization of
the Gaussian functions has been used in \cite{1994NuPhA.571..447V},
\cite{1994PhRvC..50..189V} for calculations of the structure  of light nuclei in many-cluster models.
If
we involve $N_{G}$ Gaussian functions to describe the two-cluster subsystem, we
obtain $N_{G}$ solutions of the system of equations (\ref{eq:M09}). One of
them is the ground state for the lowest orbital momentum or the lowest bound
state for larger orbital momenta, and other solutions are excited pseudo-bound
states of the nucleus represented by two-cluster configurations. In our
analysis of the Pauli resonance states, we neglect all pseudo-bound states and
account for  only the ground state. Such restriction is relevant to
the physical reality for selected nuclei and makes our analysis more transparent.

At the second stage, we find the phase shift $\delta_{J}$ of the scattering of the
third cluster on the two-cluster subsystem by solving the system of linear equations
(\ref{eq:M10}) or (\ref{eq:M20}).

Having determined the phase shifts $\delta_{J}$ as functions of the energy $E$, we
can determine the energy $E_{r}$ and the total width $\Gamma$  for an
isolated resonance state  from the equations%
\begin{equation}
\left.  \frac{d^{2}\delta_{J}}{d^{2}E}\right\vert _{E=E_{r}}=0,\quad
\Gamma=\left.  \left(  \frac{d\delta_{J}}{dE}\right)  ^{-1}\right\vert
_{E=E_{r}},
\label{eq:M27}
\end{equation}
which utilize the well-known Breit--Wigner formula.

\section{Results and discussions\label{Sec:Results}}

As was indicated above, we consider a set of light nuclei. In Table
\ref{Tab:Parameters}, we list these nuclei and present details of the model and
calculations. Here, 3C stands for a three-cluster configuration which is taken
into consideration, BC indicates binary channels which are studied. The
Minnesota potential (MP) \cite{kn:Minn_pot1} is used as a nucleon-nucleon
potential. The oscillator length $b$ is chosen to minimize the energy of the three-cluster
threshold. The exchange parameter $u$ of the MP is usually selected to
reproduce the  ground-state energy of a compound system accounted from the
lowest two- or three-body threshold.

For all nuclei (but not for $^{6}$Li) listed in Table \ref{Tab:Parameters}, we employ
only one three-cluster configuration. For $^{6}$Li, we employ two
three-cluster configurations $^{4}$He+$p+n$ and $^{3}$H+$d+p$. The first
three-cluster configuration allows us to consider the dominant binary channel
$^{4}$He+$d$ and describe a deuteron as the $p+n$ two-body system. The second
three-cluster configuration is used to study the second binary channel $^{3}%
$H+$^{3}$He and to describe the nucleus $^{3}$He which is less bound than
$^{3}$H, as the two-cluster structure $d+p$. We do not consider the alternative
three-cluster configuration $^{3}$He+$d+n$, as the configuration $^{3}$H+$d+p$
suggests a more realistic description of the $^{3}$H+$^{3}$He channel.

In the present paper, as was mentioned above, we employ the three-cluster
model, which was designed to study the interaction of three $s$-clusters. This
allows us to study the scattering of $s$-shell clusters (such as $n$, $d$, $^{3}%
$H, $^{3}$He, $^{4}$He) on $s$-shell clusters, and the scattering of $s$-shell
clusters on $p$-shell clusters, such as $^{6}$Li, $^{8}$Be.  One can see from
Table \ref{Tab:Parameters}, we selected clusters of $s$-shell  and those
$p$-shell clusters, which have zero value of the internal orbital momenta
$L_{1}$=$L_{2}$=0 (see Eqs. (\ref{eq:M01}) and (\ref{eq:M02})).  This
selection  substantially simplifies calculations and allows us to use the
single-channel approximation. As the results, the total orbital momentum $L$ 
coincides with the orbital momentum of relative motions of clusters. Besides, 
the spins $S_{1}$ and $S_{2}$ of the first and second clusters are good quantum 
numbers, even if the spin-orbit components of a nucleon-nucleon potential are involved.

We employ four Gaussian functions to obtain the energy and wave functions of
two-cluster subsystems, and 100 oscillator functions to describe the scattering of
the third cluster on the two-cluster subsystem. It was checked numerously that
such number of oscillator functions is sufficient to obtain the bound-state
energies of a compound nucleus and the scattering parameters with acceptable precision.%

%TCIMACRO{\TeXButton{B}{\begin{table}[tbp] \centering}}%
%BeginExpansion
\begin{table}[ht] \centering
%EndExpansion
\begin{ruledtabular}  
\caption{List of nuclei to be considered, their three-cluster configurations (3C), binary channels (BC), and input parameters of calculations: oscillator length $b$, exchange parameter $u$ of the  Minnesota potential \cite{kn:Minn_pot1}. \label{Tab:Parameters}}%
\begin{tabular}
[c]{cccccc}%\hline
Nucleus & 3C & BC & $b$, fm & $u$ & Source\\\hline
$^{6}$Li & $^4$He+$p+n$ & $^4$He+$d$ & 1.285 & 0.863 &
\cite{KALZIGITOV2021APP}\\
& $^3$H+$d+p$ & $^3$H+$^{3}$He &  &  & \\%\hline
$^{7}$Li & $^4$He+$d+n$ & $^4$He+$^{3}$H, $^{6}$Li$+n$ & 1.311 & 0.956 &
\cite{2009PAN....72.1450N}\\%\hline
$^{7}$Be & $^4$He+$d+p$ & $^4$He+$^{3}$He, $^{6}$Li$+p$ & 1.311 & 0.956 &
\cite{2009NuPhA.824...37V}\\%\hline
$^{8}$Be & $^4$He+$d+d$ & $^{6}$Li+$d$ & 1.311 & 0.956 & \\%\hline
$^{9}$Be & $^4$He+$^3$H+$d$ & $^{6}$Li+$^3$H & 1.285 & 0.950 & \cite{2023arXiv231013979L, 2024FBS....65...14L} 
\\%\hline
$^{10}$B & $^4$He+$^4$He+$d$ & $^{8}$Be+$d$, $^{6}$Li+$^4$He & 1.298 & 0.900 &
\cite{2014UkrJPh..59.1065N}\\%\hline
\end{tabular}
\end{ruledtabular}  
%\label{Tab:Parameters}%
%TCIMACRO{\TeXButton{E}{\end{table}}}%
%BeginExpansion
\end{table}%
%EndExpansion

To consider properties of the Pauli resonances in more details, we restrict
ourselves to the single-channel approximation. Moreover, we do not consider
mixture of states with different values of the total orbital momentum $L$ and
total spin $S$, thus in our present model $L$ and $S$ are additional quantum
numbers to the angular momentum $J$ and parity $\pi$ of a compound system. In
this paper we will not consider many-channel cases. This is a subject for our
next investigation.

\subsection{Manifestation of the Pauli resonance states \label{Sec:Manifest}}

\begin{figure}[hptb]
\begin{center}
\includegraphics[width=\textwidth]{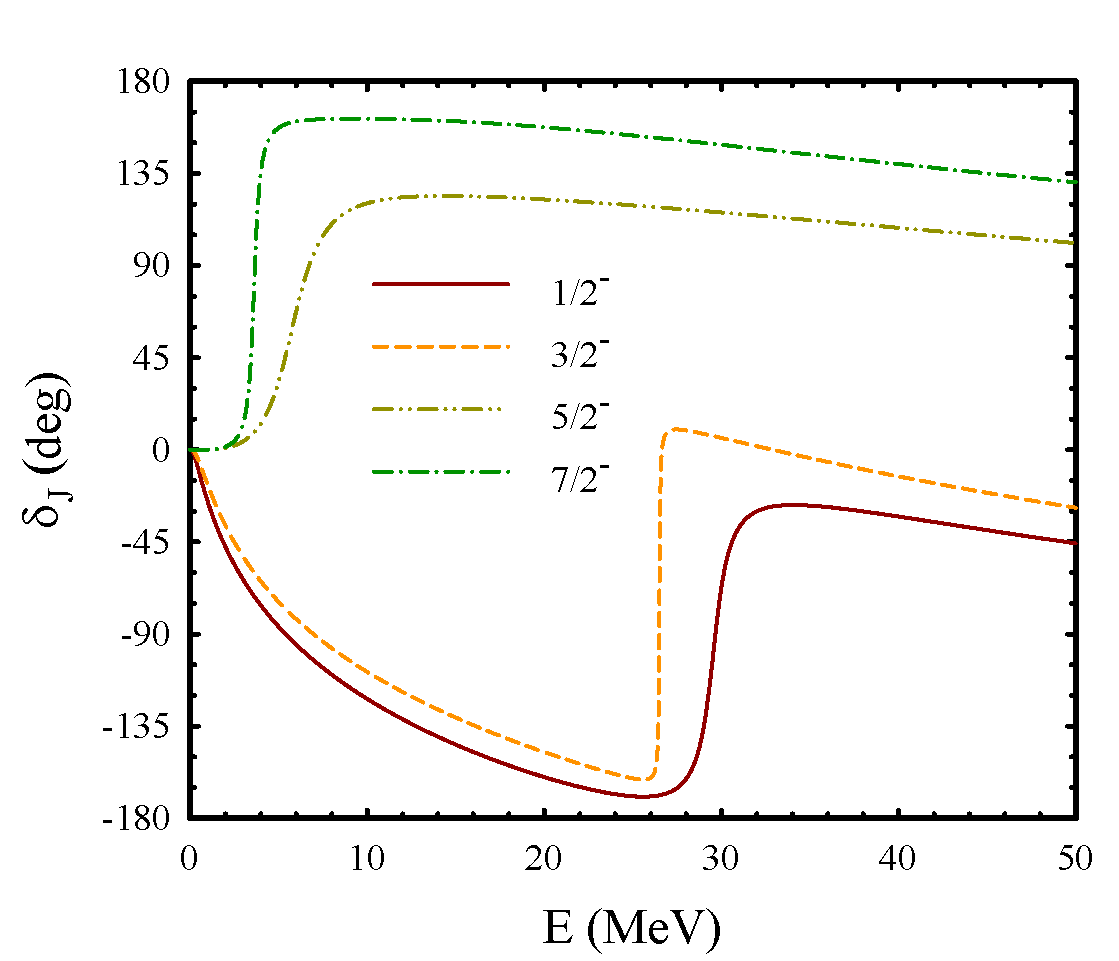}%
%{PhasesAT.jpg}%
\caption{Phase shifts of the $^4$He+$^3$H elastic scattering in the states
$J^{\pi}$ = 1/2$^{-}$, 3/2$^{-}$, 5/2$^{-}$ and 7/2$^{-}$. Results are obtained 
with the $^4$He+$d+n$ configuration.}%
\label{Fig:PhasesAT}%
\end{center}
\end{figure}

In this subsection, we show how the Pauli resonance states manifest themselves in
the continuous spectrum in the standard,  improved and advanced versions of the RGM calculations. 
For this aim, we consider phase shifts. The most
typical picture is shown in Fig. \ref{Fig:PhasesAT}, where the phase shifts are
displayed for four different $J^{\pi}$ states of the elastic $^4$He+$^3$H %$\alpha+t$ 
scattering in the $^4$He+$d+n$ calculations. These phase shifts exhibit resonance states of two different
types. The first type is the shape resonance states. They are created in the
states 7/2$^{-}$ and 5/2$^{-}$. These resonance states are formed by a
combination of a huge centrifugal barrier at the $L$=3 state and a Coulomb
barrier. The shape resonance states lie close to the $^4$He+$^3$H threshold. The
second type is the Pauli resonances which exhibit themselves in the states
3/2$^{-}$ and 1/2$^{-}$. The energies of the Pauli resonances are $E$=25.8 MeV for
3/2$^{-}$ state and $E$=29.0 MeV for 1/2$^{-}$ state. The total orbital
momentum $L$=1 is responsible for these resonance states. This means that, in
these states, the centrifugal barrier is by approximately 6 times lower than that in the
7/2$^{-}$ and 5/2$^{-}$ states. Thus, the centrifugal barrier cannot be
responsible for the Pauli resonance states. The phase shifts displayed in Fig.
\ref{Fig:PhasesAT} are similar to the phase shifts which are shown in Fig. 1
of Ref. \cite{1974PhLB...49..308C}. The improved version of the RGM has been
used in Ref. \cite{1974PhLB...49..308C}, and two different oscillator lengths
were chosen for an alpha particle and a triton.%

%BeginExpansion
%EndExpansion

\begin{figure}[hptb]
\begin{center}
\includegraphics[width=\textwidth]{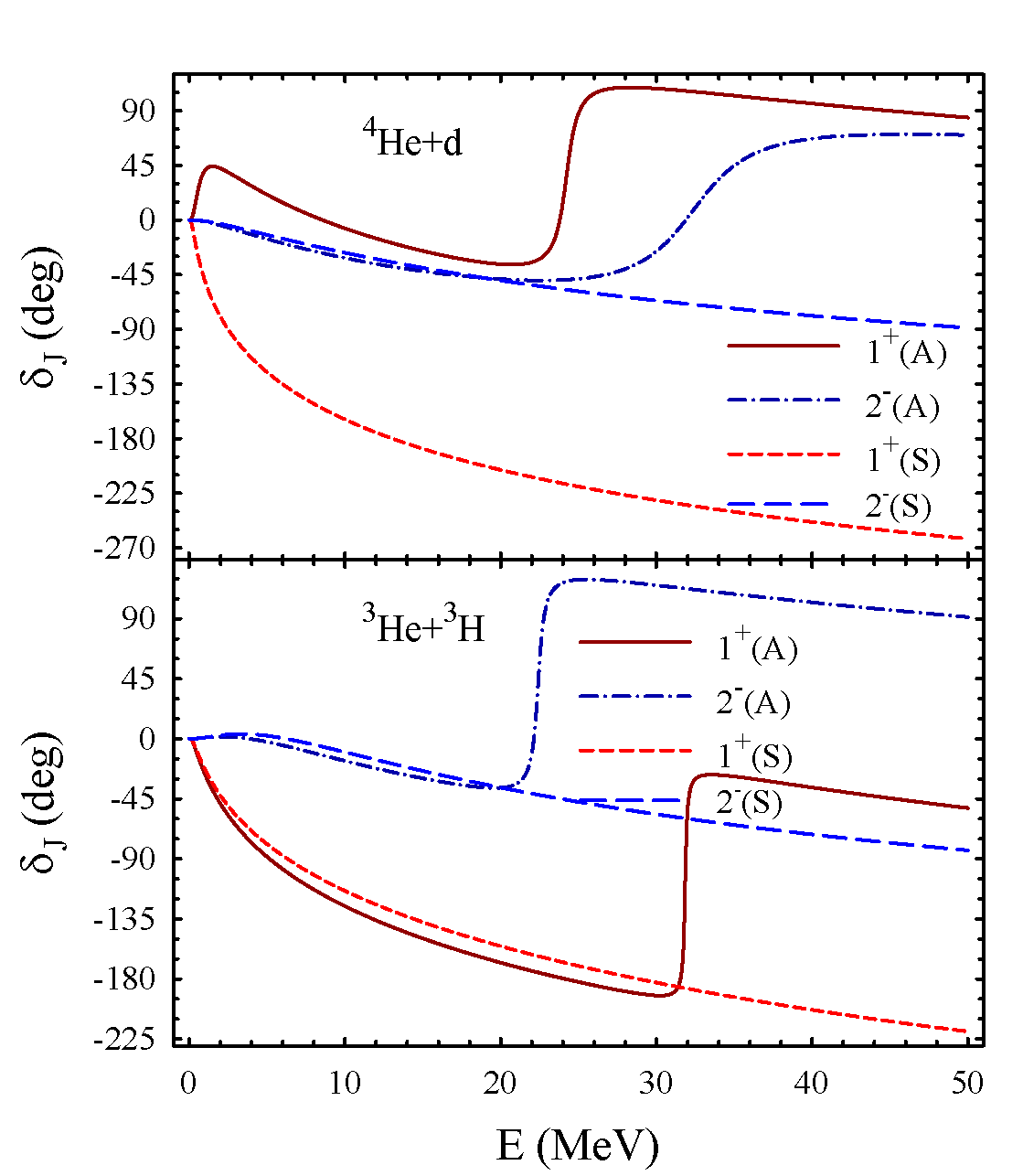}%
%{Phases6Li2Chs.jpg}%
\caption{Phase shifts of the elastic $^{4}$He+$d$ and $^{3}$He+$^3$H calculated
for the 1$^{+}$ and 2$^{-}$ states in the advanced version of RGM. Results are 
 obtained with the standard (S) and advanced (A) versions of the RGM by using 
the $^4$He+$p+n$ and $^3$H+$d+n$ configurations, respectively.}%
\label{Fig:Phases6Li2Chs}%
\end{center}
\end{figure}

Let us consider the Pauli resonance states in the 1$^{+}$ ($L$=0, $S$=1) and
2$^{-}$ ($L$=1, $S$=1) states of $^{6}$Li which exhibit themselves in the
channels $^{4}$He+$d$ and $^{3}$He+$^3$H. 
In Fig. \ref{Fig:Phases6Li2Chs} we display the phase shifts obtained in the advanced (A) and
 standard (S)  versions of the RGM. The phase shifts are
displayed for two different $J^{\pi}$ states of the elastic $^4$He+$d$ and $^3$He+$^3$H 
scattering obtained with the $^4$He+$p+n$ and $^3$H+$d+p$ configurations, respectively. As we can see, the phase shifts, obtained
in the standard versions of the RGM, are monotonic functions of the energy,
they do not exhibit the resonance behavior.
Meanwhile, the phase shifts, in the advanced versions of the RGM, exhibit resonance states  
in the 1$^{+}$ ($L$=0, $S$=1) and 2$^{-}$ ($L$=1, $S$=1) states of $^{6}$Li.
Note that at the low-energy region, the phase shifts obtained in the advanced
(A) and standard (S) versions of the RGM are close one another. However,
that is not the case for the phase shifts of the $^{4}$He+d scattering in the
$1^{+}$ states (i.e., in the state with $L$=0, and $S$=1). Such a difference
of the phase shift behavior is determined by the position of the $^{6}$Li
ground state in these two models. With the input parameters selected, the
ground state is slightly bound in the advanced version, and in the standard
version it is a pseudo-bound state.

We recall that the wave
functions of a deuteron and $^{3}$He are obtained in the two-cluster (two-body)
approximations as $p+n$ and $d+p$, respectively. Such advanced description of
a deuteron and a triton stipulate the appearance of the Pauli resonance states shown
in Fig. \ref{Fig:Phases6Li2Chs}. Only one Pauli resonance state is found in
each channel. The energies and widths of these resonances depend on the total
orbital momentum $J$. %
One notices that there are two resonance states in the channel $^{4}$He+d with
$J^{\pi}$=1$^{+}$, and one of them is the Pauli resonance state with the energy
$E$=24.2 MeV. The second one with the energy $E$=0.257 MeV and the width $\Gamma
$=0.226 is the shape resonance state. The composition of the attractive nuclear and
repulsive Coulomb interactions in the three-cluster system $^{4}$He+$p+n$
created a favorable condition for creating the low-energy resonance. It is
shown in Ref.  \cite{KALZIGITOV2021APP}, that this resonance state is
transformed into the bound state in the four-channel approximation.

%
%BeginExpansion
%EndExpansion

\begin{figure}[hptb]
\begin{center}
\includegraphics[width=\textwidth]{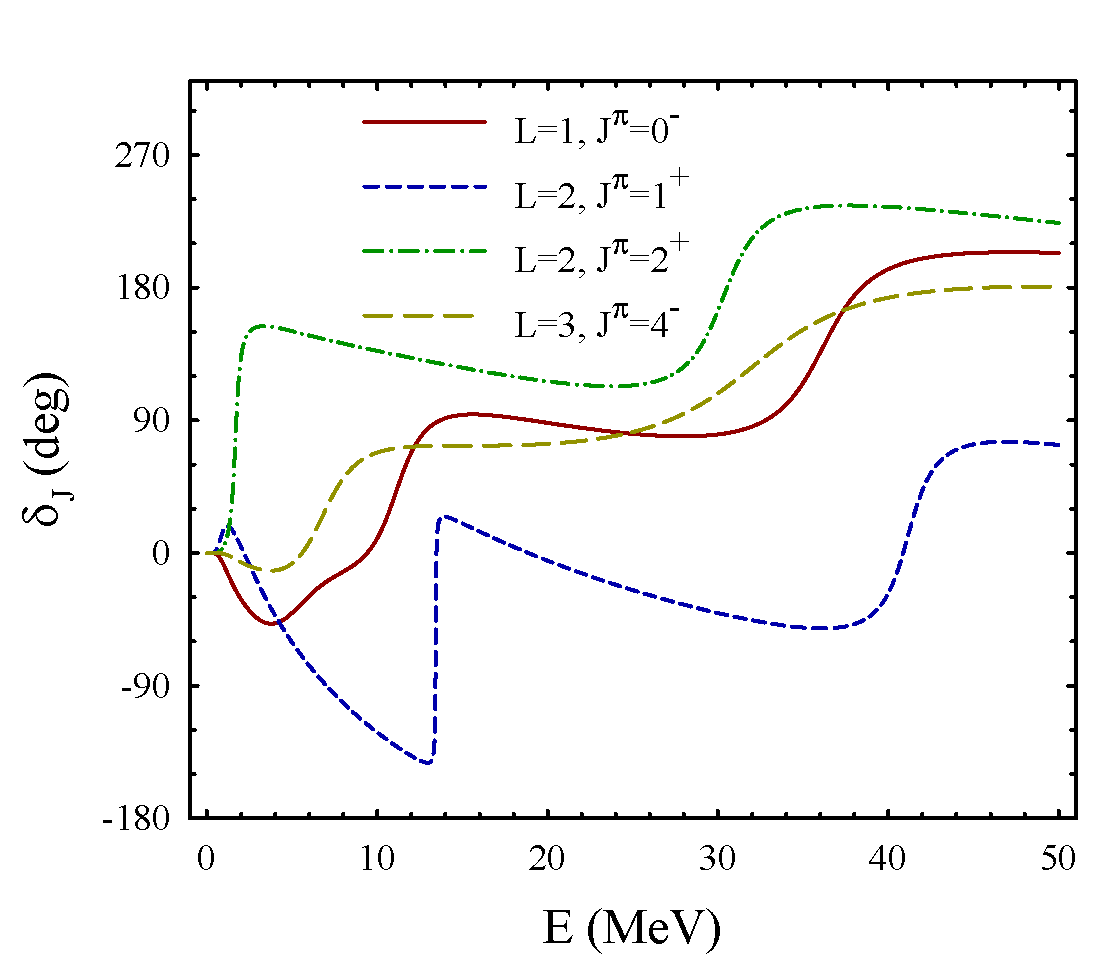}%
%{../Phases6LiA.jpg}%
\caption{Phase shifts of the elastic $^{6}$Li+$^4$He scattering as a
function of the energy $E$. Results are obtained by the $^4$He+$^4$He+$d$ calculation.}%
\label{Fig:Phases6LiA}%
\end{center}
\end{figure}

Another example of the Pauli resonance manifestation is shown in Fig.
\ref{Fig:Phases6LiA} for the $^{6}$Li+$^4$He scattering. This case
demonstrates that the two-cluster system may have two Pauli resonance states, they
are located at energy range 10$\leq E\leq$45 MeV. A sharp growing of the
1$^{+}$ phase shifts around 13.4 MeV indicates that there is very narrow
resonance state with the widths $\Gamma$=56 keV. Other Pauli resonances are
significantly wider.
%
%BeginExpansion
%EndExpansion

\begin{figure}[hptb]
\begin{center}
\includegraphics[width=\textwidth]{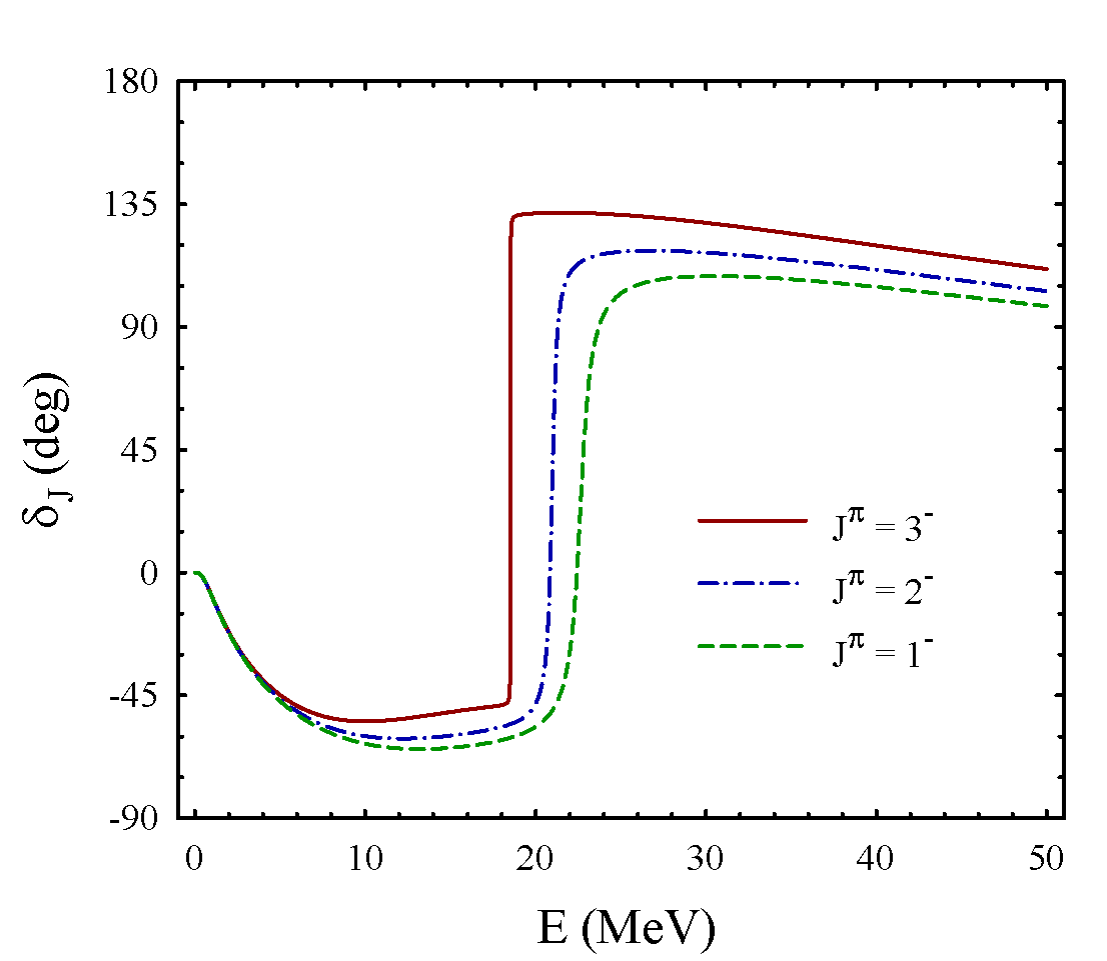}%
%{../Phases8BeDL1S2.jpg}%
\caption{Phase shifts of the elastic $^{6}$Li+$d$ scattering as a function of the
energy. They are obtained by the $^4$He+$d+d$ calculation for $L$=1, $S$=2, 
and three values of the total angular momentum $J$.}%
\label{Fig:Phases8BeDL1S2}%
\end{center}
\end{figure}

In Fig. \ref{Fig:Phases8BeDL1S2}, we show the phase shifts of the elastic $^{6}%
$Li+$d$ scattering with the total orbital momentum $L$=1 and total spin $S$=2.
Due to the spin-orbit potential, we obtain phase shifts for three states with
the total angular momenta $J^{\pi}$ =3$^{-}$, 2$^{-}$, and 1$^{-}$. Thus, the 
difference in behavior of phase shifts is totally originated from the
spin-orbit potential. Figure \ref{Fig:Phases8BeDL1S2} demonstrates that the energies
and widths of the Pauli resonance states depend on the spin-orbit components
of the nucleon-nucleon interaction. Indeed, we obtained that $E$=22.636 MeV,
 $\Gamma$=0.952 MeV for $J$=1$^{-}$, $E$=20.981 MeV,  $\Gamma$=0.402 MeV for
$J$=2$^{-}$, $E$=18.523 MeV, $ and \Gamma$=0.008 MeV for $J$=3$^{-}$. We see that the
spin-orbit potential substantially changes the energies and widths of the Pauli
resonance states.%

%BeginExpansion
%EndExpansion

\begin{figure}[hptb]
\begin{center}
\includegraphics[width=\textwidth]{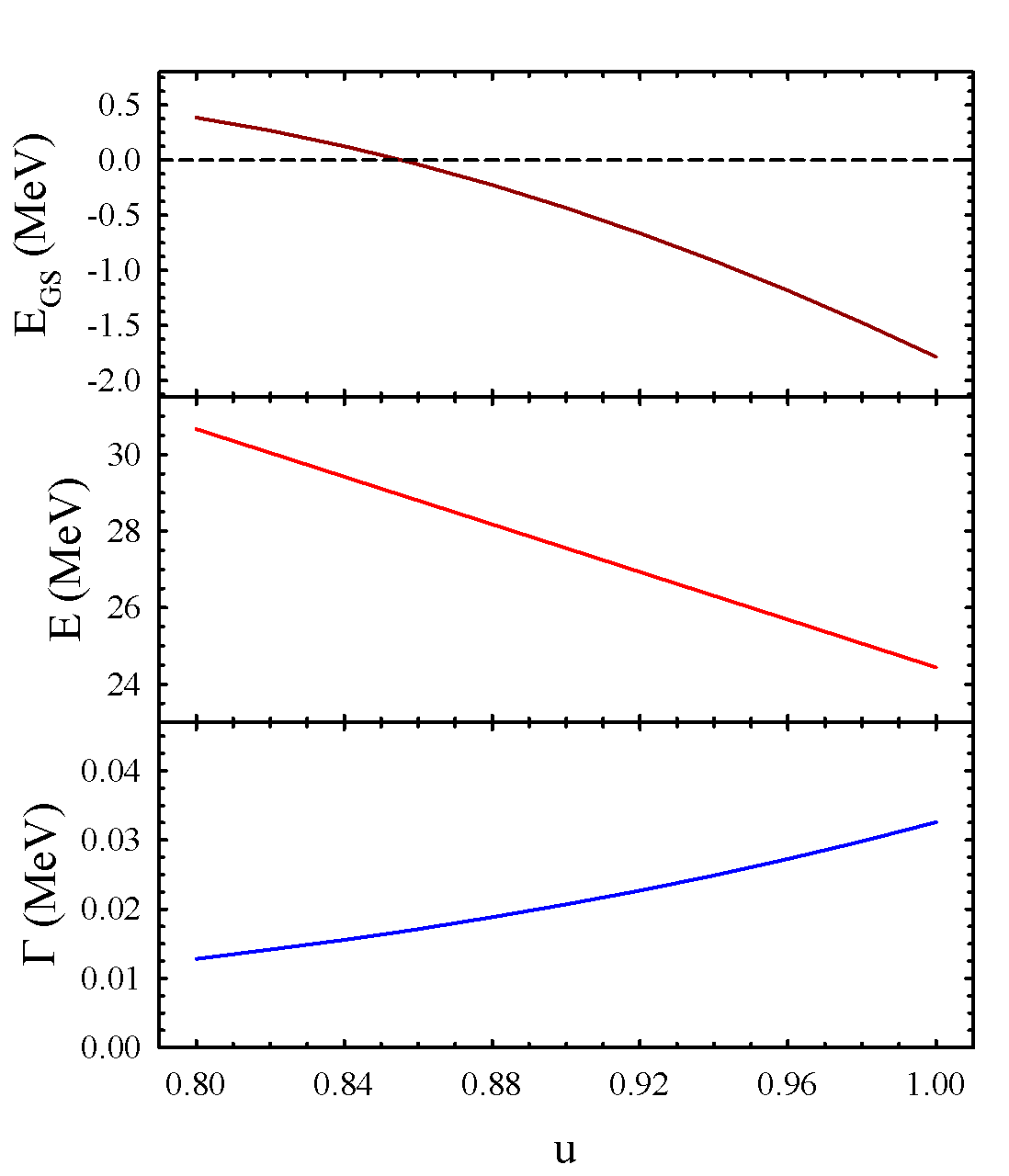}%
%{../Paramts7Li32MvsU.jpg}%
\caption{Dependence of the ground-state energy $E_{GS}$ of $^{7}$Li, the energy
and width of the Pauli resonance state in the 3/2$^{-}$ state of the
$^4$He+$^3$H channel on the exchange parameter $u$ of the MP. Calculations are performed with 
the three-cluster configuration $^4$He+$d+n$.}%
\label{Fig:Params7Li32MvsU}%
\end{center}
\end{figure}

Let us consider how the central part of the MP affects the energies and widths of the
Pauli resonance states. This can be done by varying the exchange parameter $u$
of that potential. This parameter affects the interaction of nucleons in odd
states, as well as the cluster-cluster interaction. The smaller is $u$, the
smaller is the interaction of clusters. When the parameter $u$ approaches 
unity, the interaction of clusters  increases. The influence of the parameter $u$
variation is carried out for the 3/2$^{-}$ state of $^{7}$Li considered as
two-cluster system $^4$He+$^3$H. Results of the variation of $u$ are demonstrated for
the ground-state energy $E_{GS}$ and for the energy and width of the Pauli
resonance. On can see in Fig. \ref{Fig:Params7Li32MvsU} that the parameter $u$
changes the energy of the ground state. Moreover, when $u<$%
0.86, the nucleus $^{7}$Li has no bound state. By varying the parameter $u$ from
0.86 to 1, we change the ground-state energy from -0.038 to -1.78 MeV.
However, the variation of $u$ from 0.8 to 1 reduces significantly the energy of the
Pauli resonance from 30.67 to 24.44 MeV. Such a variation of $u$ slightly
changes the width of the Pauli resonance state from 13 to 33 keV.

Note an unusual feature of the Pauli resonance  displayed in Fig. 
\ref{Fig:Params7Li32MvsU}: the width of the resonance is decreasing with 
incising of the exchange parameter $u$. For shape resonances states is usually 
observed another feature, both energy and width are decreasing with increasing 
of the parameter $u$.

%BeginExpansion
%EndExpansion

\subsection{Special case for $^4$He+$d$ system}

Taking into account peculiarities of our model, we decided to carry out an
additional investigation of the $^4$He+$d$ system. In this specific case, our
model allows us to realize not only the standard and advanced, but also improved
version of the RGM. If we take only one Gaussian function in the expansion of
the deuteron wave function (i.e. to describe relative motion of a 
structureless proton and neutron) and select the parameters $b_{0}$ (see Eqs. (\ref{eq:M04}%
) and (\ref{eq:M22})) to minimize the bound-state energy of a deuteron, we realize therefore the
improved version of the RGM. One Gaussian function with the optimal value of
$b_{0}=$1.512 fm creates a bound state of a deuteron with the energy $E$ =-0.132 MeV,
while four Gaussian functions with optimal values of $b_{0}$ and $q$ (see Eq. (\ref{eq:M22})) 
generate the deuteron bound state with the energy $E$ =-2.020 MeV.

\begin{figure}%[ht]
\begin{center}
\includegraphics[width=\textwidth]{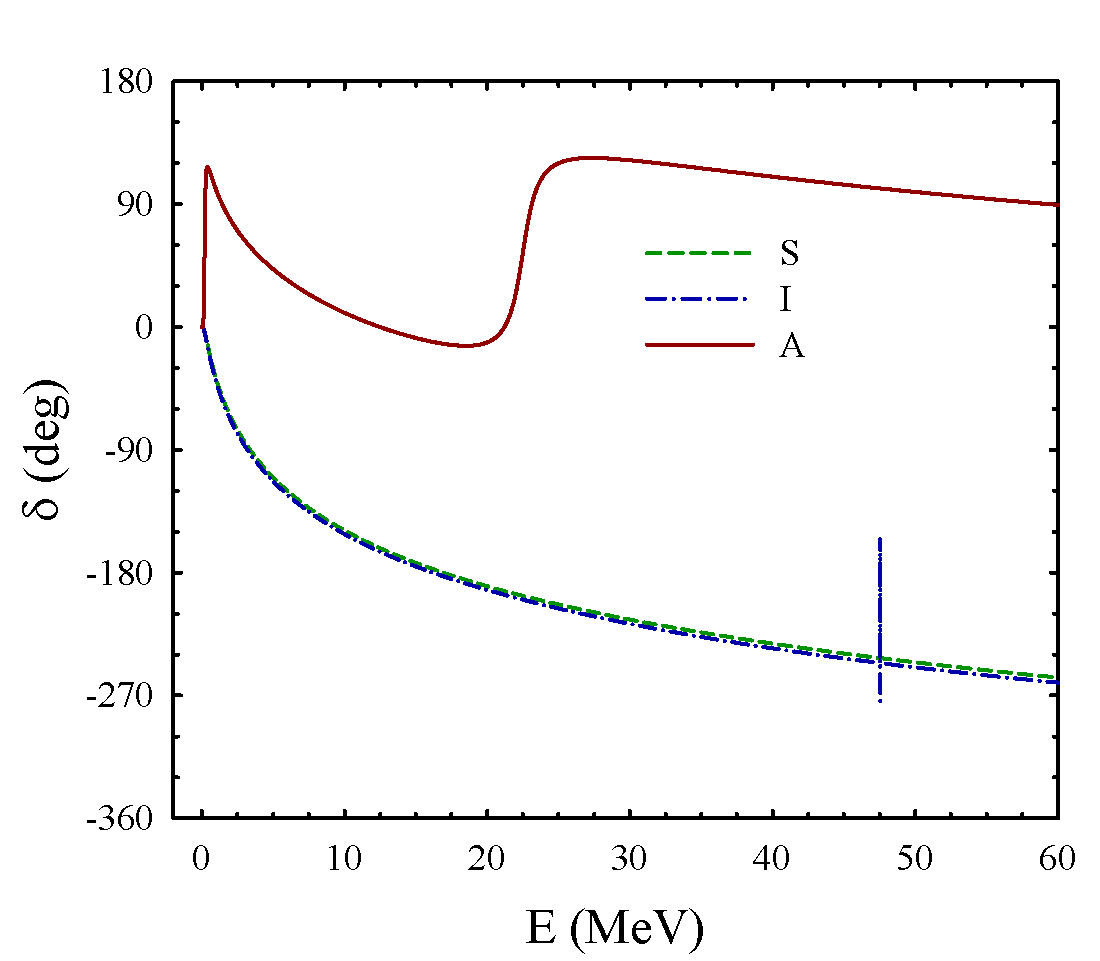}%
%{PhasesAD1P3App.jpg}%
\caption{Phase shifts of the elastic $^4$He+$d$ scattering in the state $L=0$,
$S=1$, $J^{\pi}=$1$^{+}$, obtained in three different approximations of the
RGM and with the $^4$He+$p+n$ calculation.}%
\label{Fig:PhasesAD1P3App}%
\end{center}
\end{figure}

To locate the Pauli resonance state in the approximation in the energy region
below 50 MeV, we have to change the exchange parameter $u$ and take $u$=1.0.
In Fig. \ref{Fig:PhasesAD1P3App}, we show the phase shift of the $^4$He+$d$
scattering in the $L$=0, $S$=1 $J^{\pi}=1^{+}$ state obtained in three different
approximations. The standard version of the RGM does not generate the Pauli
resonance state in this case. The Pauli resonance states appear in the
improved (I) and advanced (A) versions. The parameters of the Pauli resonance
states substantially depend on the wave function describing the internal structure
of a deuteron. A more realistic wave function significantly increases the width of the
Pauli resonance (from $\Gamma=$0.001 MeV to $\Gamma=$1.718 MeV), and
dramatically changes the energy (from $E=$47.55 MeV to $E=$22.49 MeV) of the
resonance state in the state $J^{\pi}=1^{+}$.%

%BeginExpansion
%EndExpansion

\subsection{Main properties of the Pauli resonance states}

In Tables \ref{Tab:PauliResons1} and \ref{Tab:PauliResons2} we collect 
information on the parameters of
Pauli resonance states detected in nuclei under consideration. It is detected
28 Pauli resonance states. The energy of resonance states is reckoned from the
threshold of the channel indicated in the column "Channel" of Tables
\ref{Tab:PauliResons1} and \ref{Tab:PauliResons2} and varies from 11 to 46 MeV. 
There are 10 narrow resonance states with $\Gamma<$%
1 MeV, six of them are very narrow resonance states with the width $\Gamma<$%
0.1 MeV. The rest 18 resonance states are wider ones; their widths exceed 1
MeV. One can see that, in most cases, the two-cluster system with fixed quantum
numbers $L$, $S$ and $J^{\pi}$ has only one Pauli resonance state. However,
there are some cases, where two Pauli resonance states are observed. The
larger the energy of a resonance state, the larger is the total width. The energy
of the second resonance state in $^{7}$Li and $^{8}$Be is approximately by 15 MeV
larger than the energy of the first resonance state. In $^{10}$B, the energy
difference  is more than 25 MeV.%

%TCIMACRO{\TeXButton{B}{\begin{table}[tbp] \centering}}%
%BeginExpansion
\begin{table}[ht] 
\centering
%EndExpansion
\begin{ruledtabular}  
\caption{Parameters of the Pauli resonance states in $^6$Li, $^7$Li 
and $^{10}$B. \label{Tab:PauliResons1}}%
\begin{tabular}
[c]{ccccccc}%\hline
Nucleus & Channel & $L$ & $S$ & $J^{\pi}$ & $E$, MeV & $\Gamma$, MeV\\\hline
$^{6}$Li & $^4$He+$d$ & 0 & 1 & 1$^{+}$ & 24.218 & 1.165\\
&  & 1 & 1 & 2$^{-}$ & 32.370 & 6.755\\
& $^{3}$He+$^3$H & 0 & 1 & 1$^{+}$ & 31.844 & 0.209\\
&  & 1 & 1 & 2$^{-}$ & 22.403 & 0.618\\\hline
$^{7}$Li & $^4$He+$^3$H & 1 & 1/2 & 1/2$^{-}$ & 29.002 & 2.144\\
&  & 1 & 1/2 & 3/2$^{-}$ & 25.810 & 0.027\\
&  & 0 & 1/2 & 1/2$^{+}$ & 20.148 & 2.589\\
&  & 0 & 1/2 & 1/2$^{+}$ & 34.444 & 4.702\\
& $^{6}$Li$+n$ & 0 & 1/2 & 1/2$^{+}$ & 12.863 & 3.332\\
&  & 0 & 3/2 & 3/2$^{+}$ & 18.895 & 0.196\\\hline
$^{10}$B & $^{6}$Li+$^4$He & 1 & 1 & 0$^{-}$ & 11.090 & 3.198\\
&  & 1 & 1 & 0$^{-}$ & 35.834 & 4.600\\
&  & 1 & 1 & 1$^{-}$ & 11.098 & 3.424\\
&  & 1 & 1 & 1$^{-}$ & 36.167 & 5.105\\
&  & 0 & 1 & 1$^{+}$ & 13.427 & 0.056\\
&  & 0 & 1 & 1$^{+}$ & 41.144 & 2.751\\%\hline
\end{tabular}
\end{ruledtabular}  
%\label{Tab:PauliResons1}%
%TCIMACRO{\TeXButton{E}{\end{table}}}%
%BeginExpansion
\end{table}%
%EndExpansion

In Table \ref{Tab:PauliResons2}, we present the parameters of the Pauli resonance
states obtained in different states for the $^{6}$Li+$d$ and $^{7}$Li+$d$ scattering.%

%TCIMACRO{\TeXButton{B}{\begin{table}[tbp] \centering}}%
%BeginExpansion
\begin{table}[ht] \centering
%EndExpansion
\begin{ruledtabular}  
\caption{Energies and widths of the Pauli resonance states in $^8$Be 
and $^9$Be observed in the channels $^6$Li+$d$ and $^7$Li+$d$, 
respectively. \label{Tab:PauliResons2}}%
\begin{tabular}
[c]{ccccccc}%\hline
Nucleus & Channel & $L$ & $S$ & $J^{\pi}$ & $E$, MeV & $\Gamma$, MeV\\\hline
$^{8}$Be & $^{6}$Li+$d$ & 0 & 0 & $0^{+}$ & 17.233 & 3.553\\
&  & 0 & 1 & $1^{+}$ & 14.989 & 1.011\\
&  & 0 & 1 & $1^{+}$ & 25.724 & 4.628\\
&  & 0 & 2 & $2^{+}$ & 20.656 & 0.008\\
&  & 1 & 0 & $1^{-}$ & 18.253 & 0.058\\
&  & 1 & 1 & $2^{-}$ & 45.555 & 6.097\\
&  & 1 & 1 & $2^{-}$ & 18.523 & 0.008\\
&  & 1 & 2 & $3^{-}$ & 18.531 & 0.013\\
&  & 1 & 2 & $2^{-}$ & 20.981 & 0.402\\\hline
$^{9}$Be & $^{7}$Li+$d$ & 1 & 1/2 & 1/2$^{-}$ & 13.733 & 1.003\\
&  & 0 & 1/2 & 1/2$^{+}$ & 15.717 & 5.796\\
&  & 0 & 1/2 & 1/2$^{+}$ & 27.958 & 1.836\\%\hline
\end{tabular}
\end{ruledtabular}  
%\label{Tab:PauliResons2}%
%TCIMACRO{\TeXButton{E}{\end{table}}}%
%BeginExpansion
\end{table}%
%EndExpansion

\begin{figure}[ht]
\begin{center}
\includegraphics[width=\textwidth]{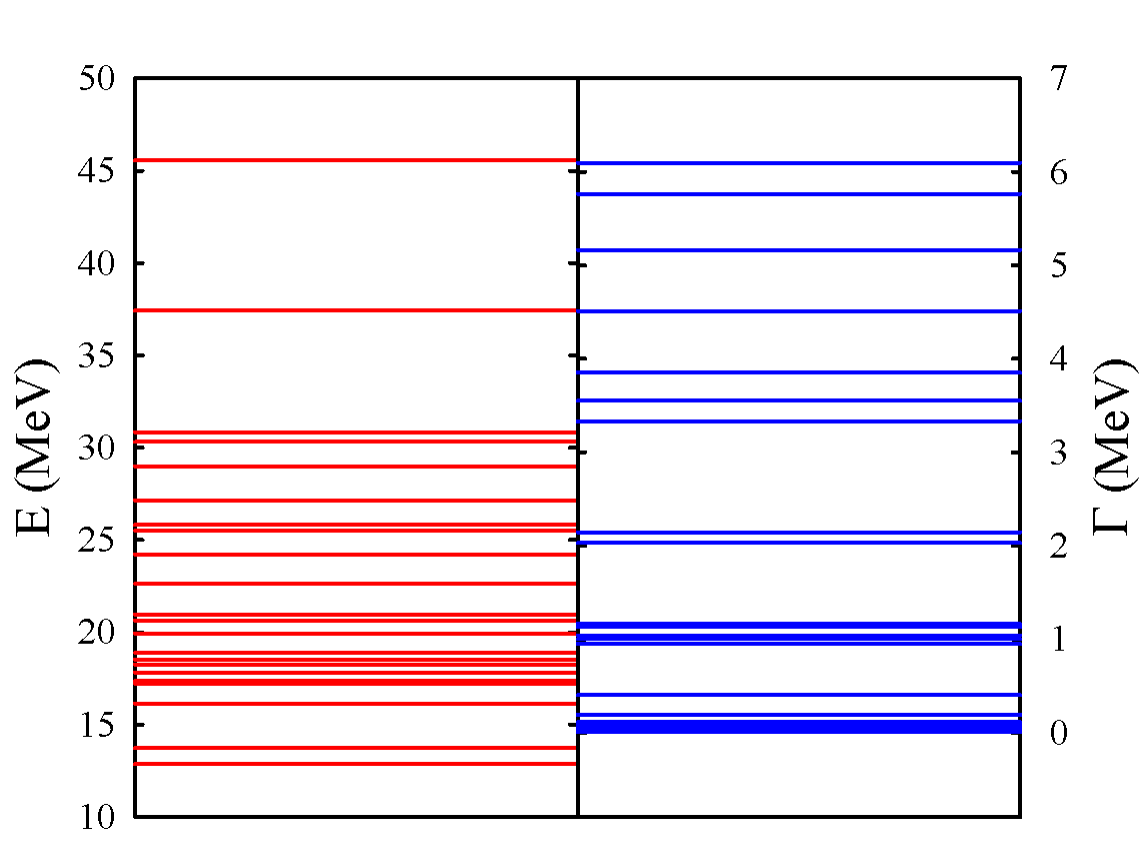}%
%{DensityEandG.jpg}%
\caption{Density of resonance state energies (left panel) and widths (right
panel) of all determined Pauli resonance states.}%
\label{Fig:DensityEandG}%
\end{center}
\end{figure}

By analyzing the results presented in Tables \ref{Tab:PauliResons1} and
\ref{Tab:PauliResons2}, we came to the conclusion that the Pauli resonance states
in light nuclei have energy more than 11 MeV, and their widths are mainly large
($\Gamma>$0.9 MeV). However, a few very narrow resonance states were found. The
most populated area of resonance states lies in the interval 16$<E<$21 MeV,
 as it is demonstrated in Fig. \ref{Fig:DensityEandG}, left panel. Two
dense area of widths of resonance states are located in the intervals 0.008
$<\Gamma<$0.22 MeV and 0.9$<\Gamma<$1.2 MeV (Fig. \ref{Fig:DensityEandG},
right panel). In many cases, only one Pauli resonance state appears in a
binary channel. We also determined several cases with two resonance states.%

%BeginExpansion
%EndExpansion

\begin{figure}[ht]
\begin{center}
\includegraphics[width=\textwidth]{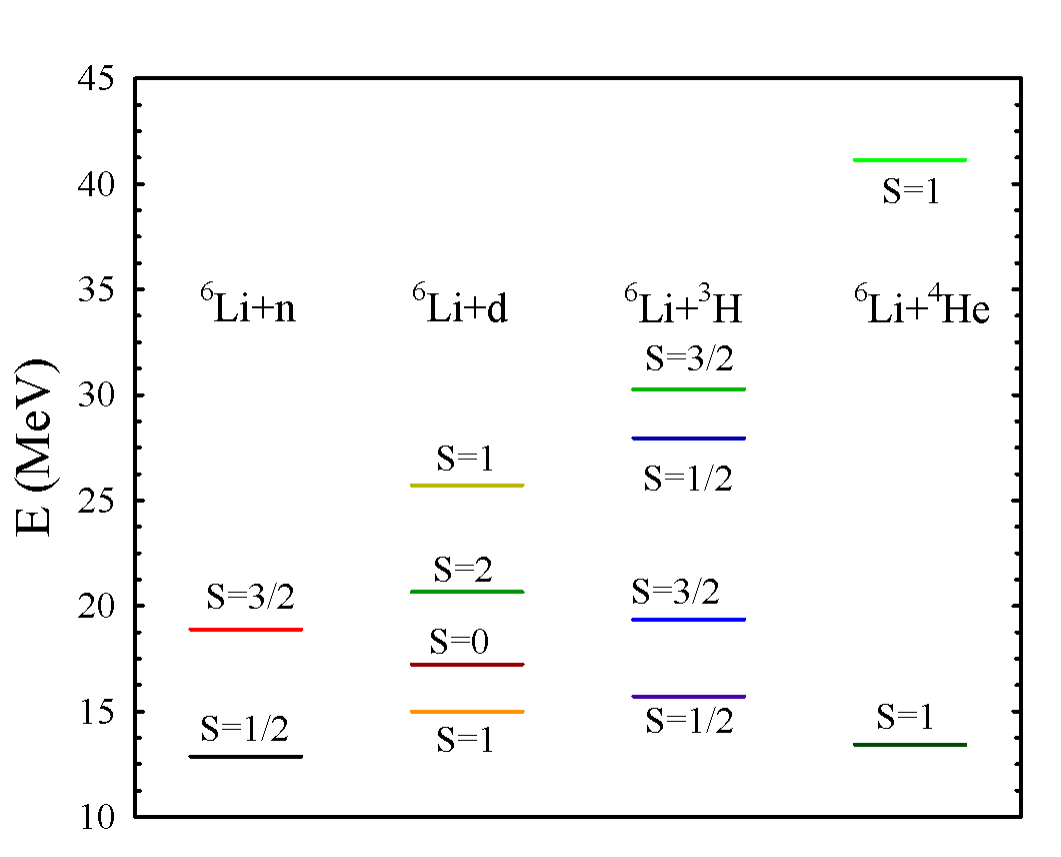}%
%{SpectrPauliResons6LiA2.jpg}%
\caption{Spectrum of positive parity Pauli resonances for $L$=0 state of nuclei $^{7}$Li,
$^{8}$Be, $^{9}$Be, $^{10}$B detected in the channels $^{6}$Li+$n$, $^{6}%
$Li+$d$, $^{6}$Li+$^3$H and $^{6}$Li+$^4$He, correspondingly.}%
\label{Fig:SpectrPR6LiA2}%
\end{center}
\end{figure}

In Fig. \ref{Fig:SpectrPR6LiA2}, we display the spectrum of Pauli resonances of
positive parity states with the total orbital momentum $L=0$. These resonance
states emerge in nuclei $^{7}$Li, $^{8}$Be, $^{9}$Be and $^{10}$B with
clusterization $^{6}$Li+$A_{2}$, where $A_{2}$\ stands for a neutron, deuteron,
triton and alpha particle. This Figure shows that the energies of the first
Pauli resonance states are quite close for all nuclei. It also shows that
there are two Pauli resonance states in the channel $^{6}$Li$+d$ with the
total spin $S$=1 and in the channel $^{6}$Li+$^3$H with the total spins $S$=1/2
and $S$=3/2. One can see that the larger the second cluster, the larger is
the energy of the highest Pauli resonance state. Indeed, it grows from 19
MeV in $^{6}$Li$+n$ channel to 41 MeV in the channel $^{6}$Li+$^4$He.%

%BeginExpansion
%EndExpansion

\begin{figure}[ht]
\begin{center}
\includegraphics[width=\textwidth]{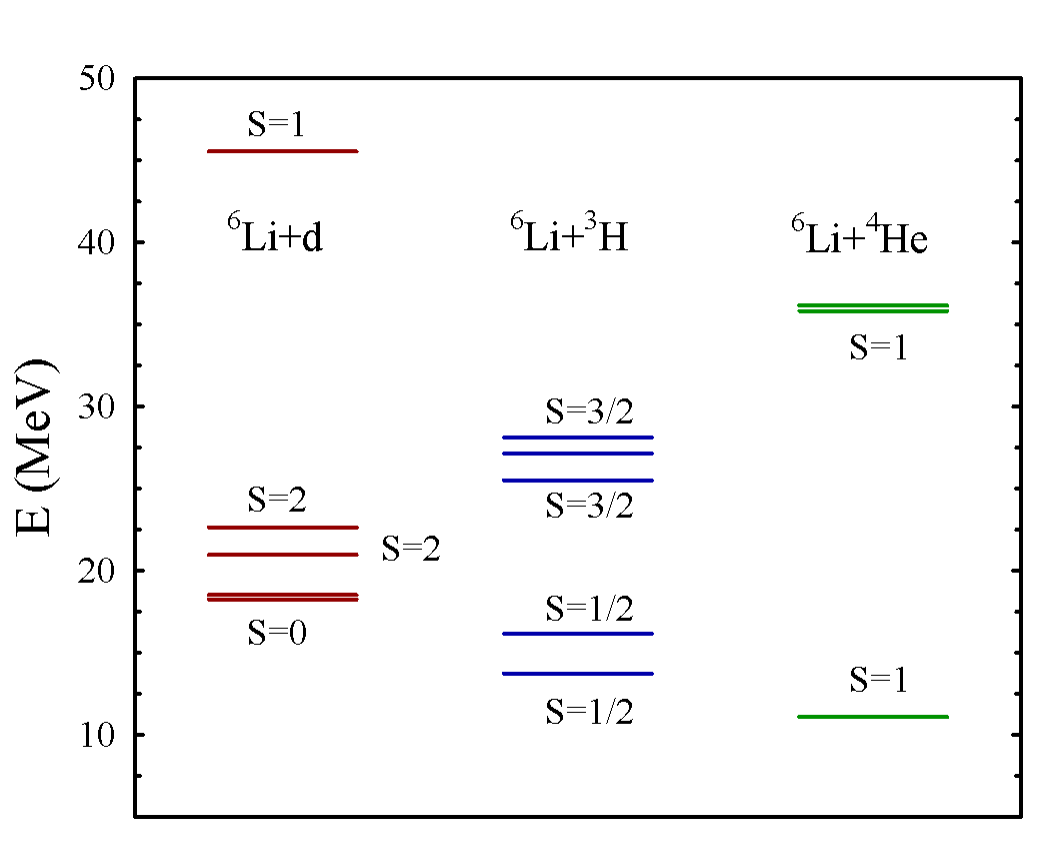}%
%{SpectrPauliResons6LiA2L1.jpg}%
\caption{Spectrum of the Pauli resonance states of the negative parity in
$^{8}$Be, $^{9}$Be and $^{10}$B created in the state with the total orbital
momentum $L$=1.}%
\label{Fig:SpectrPR6LiA2L1}%
\end{center}
\end{figure}

The Pauli resonance states of negative parity created in the channel $^{6}%
$Li+$A_{2}$ ($A_{2}$ =$d$, $^3$H, $^4$He) with the total orbital momentum
$L$=1 are shown in Fig. \ref{Fig:SpectrPR6LiA2L1}. We found no
resonance state in the channel $^{6}$Li+$n$. In this channel, the Pauli
resonance states do not appear neither in states with total spin $S$=1/2 nor
in the states $S$=3/2. Five Pauli resonance states are found in $^{8}$Be and
$^{9}$Be, and four resonances are detected in $^{10}$B. Fig.
\ref{Fig:SpectrPR6LiA2L1} shows that the energy of the lowest Pauli resonance
state decreases, as the mass of the "projectile" $A_{2}$ increases. It is
interesting tendency, as the Coulomb repulsion between $^{6}$Li and $A_{2}$ is
increases with the mass of the second cluster $A_{2}$. It is necessary
to underline that the spin-orbit interaction plays an important role in all
cases where both the total orbital momentum $L$ and total spin $S$ do not equal
to zero.%

%BeginExpansion
%EndExpansion

\subsubsection{Birth of the Pauli resonance state}

To detect the Pauli and shape resonance states, we analyzed the behavior of phase
shifts as functions of the energy. The rapid growth of the phase shift was considered as
the signal of a resonance state. There is another way for detecting the resonance
states of both types. This way is applicable for any method which involves a
square integrable basis of functions. Unfortunately, this method works for
relatively narrow resonance states. The narrow resonance states can be
detected by calculating the eigenspectrum of a Hamiltonian with different numbers
of basis functions. By displaying the eigenenergies as functions of the number of
basis functions (we denote them as $N_{O}$) involved in calculations, a
resonance state will display itself as a plateau or/and as an avoid crossing.
The energy of a plateau is the energy of a resonance state. Such way of detecting the
resonance states is an essential element of the stabilization method (Ref.
\cite{1970PhRvA...1.1109H}) and the complex scaling method (see definitions of the
method and its recent progress in applications to many-cluster systems in
Refs. \cite{2014PrPNP..79....1M, 2012PThPS.192....1H,
 2020PTEP.2020lA101M}).

\begin{figure}[ht]
\begin{center}
\includegraphics[width=\textwidth]{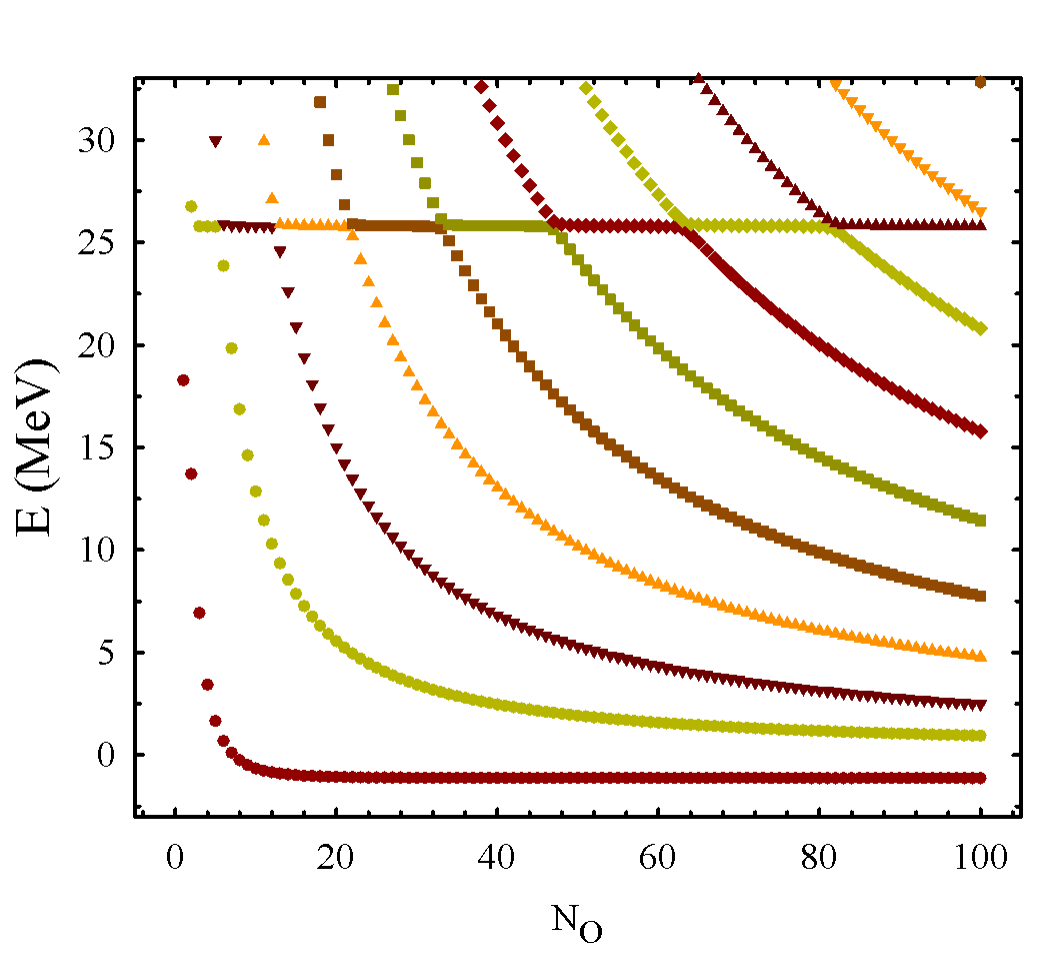}%
%{Spectr32M7LiTvsN.jpg}%
\caption{Spectrum of the 3/2$^{-}$ states in $^{7}$Li as a function of the
number of oscillator functions $N_{O}$ involved in calculations. Calculations are performed with 
the three-cluster configuration $^4$He+$d+n$.}%
\label{Fig:Spectr32M7LiTvsN}%
\end{center}
\end{figure}

\begin{figure}[ht]
\begin{center}
\includegraphics[width=\textwidth]{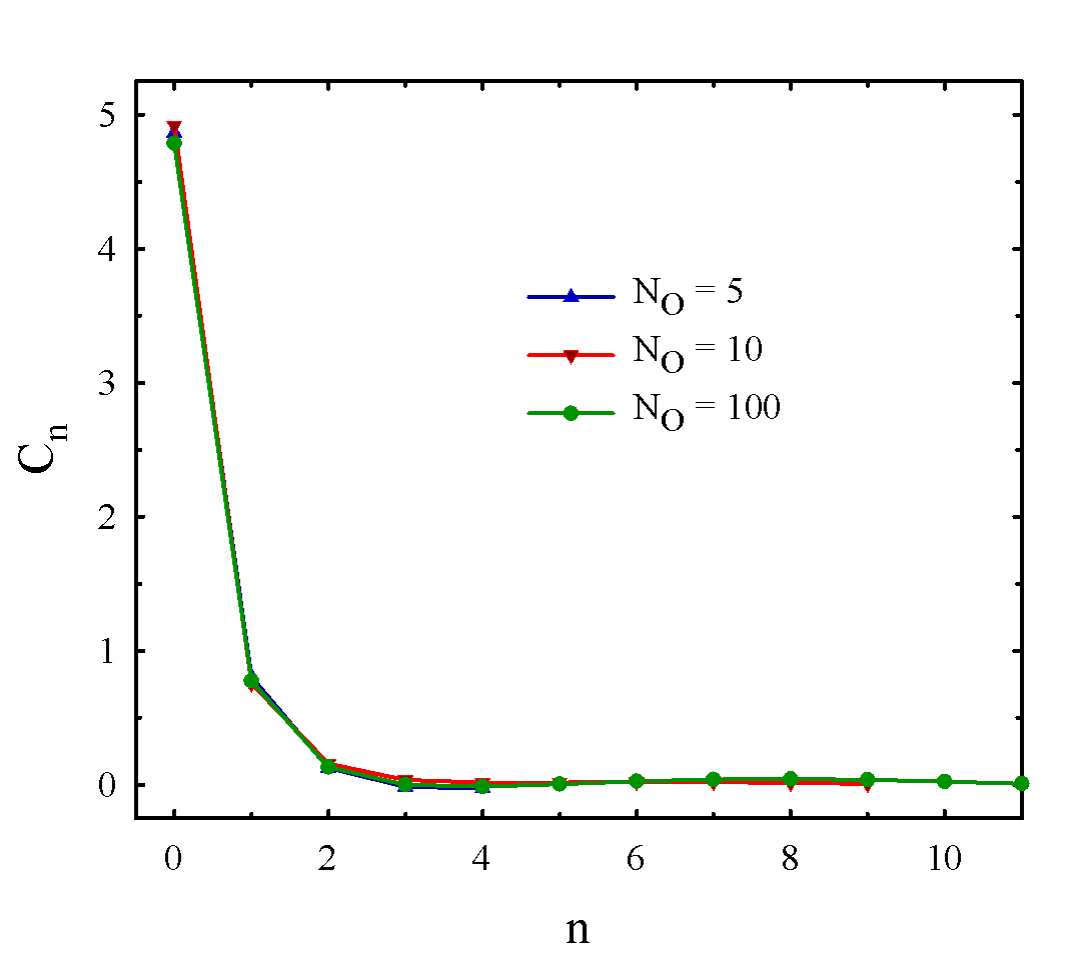}%
%{WaveFunsO32M7LiTConv.jpg}%
\caption{Convergence of the wave function of the narrow 3/2$^{-}$ Pauli resonance state
in $^{7}$Li in the channel $^4$He+$^3$H. Calculations are performed with 
the three-cluster configuration $^4$He+$d+n$.}%
\label{Fig:WaveFunsO32M7LiTConv}%
\end{center}
\end{figure}

In Fig. \ref{Fig:Spectr32M7LiTvsN}, we show the dependence of the eigenenergies of the
3/2$^{-}$ state in $^{7}$Li=$^4$He+$^3$H on the number of oscillator
functions $N_{0}$ used in calculations. We gradually change the number of
oscillator functions from 1 to 100. One can see that it is necessary to use at least three
oscillator functions to create a plateau or, in other words, to obtain the
eigenvalue with the energy which is very close to the energy of a resonance
state. Such a plateau unambiguously indicates the presence of a narrow resonance
state. This result is naturally consistent with the results of
phase shift calculations. Besides, the wave functions of the resonance states
obtained with 5, 10, and 100 oscillator functions are very close to one another
 in the region of small values of $n$, as it demonstrated in Fig.
\ref{Fig:WaveFunsO32M7LiTConv}. It proves that the narrow 3/2$^{-}$ resonance
state is formed by oscillator functions with very small values of $n$.%

%BeginExpansion
%EndExpansion
%

%BeginExpansion
%EndExpansion

\subsection{Peculiarities of the Pauli resonance states \label{Sec:PropertPR}}

Let us consider peculiarities of the wave functions of Pauli resonance
state. The analysis of wave functions will allow us to understand the nature
of the Pauli resonances. The wave functions of resonant and nonresonant states
are considered in the oscillator and coordinate representations.

\begin{figure}[ht]
\begin{center}
\includegraphics[width=\textwidth]{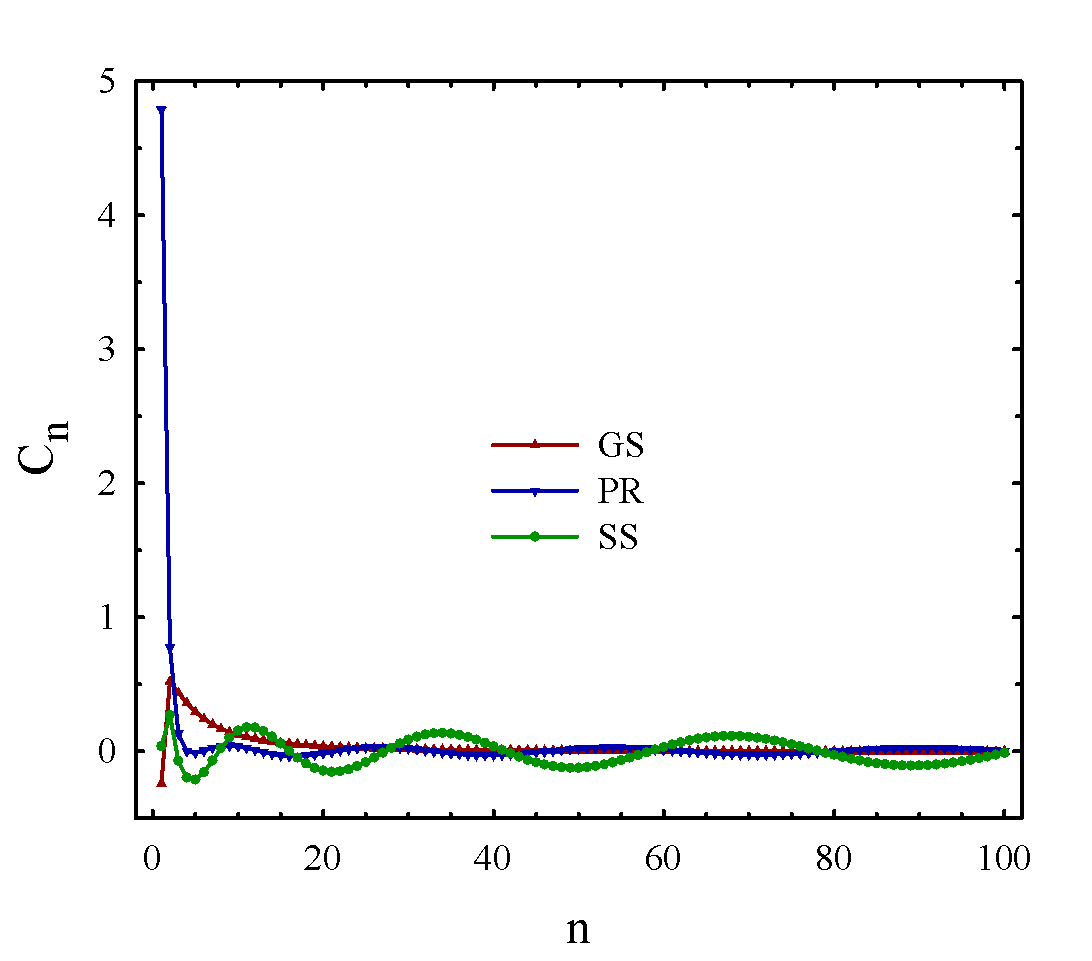}%
%{WaveFunsO32M7LiT.jpg}%
\caption{Wave functions in the oscillator representation of the ground state (GS), 
Pauli resonance state (PR) and scattering state (SS) in the 3/2$^-$ state of $^7$Li. 
Results are obtained with the $^4$He+$d+n$ calculations.}%
\label{Fig:WaveFunsO32M7LiT}%
\end{center}
\end{figure}

In Fig. \ref{Fig:WaveFunsO32M7LiT}, we show three wave functions of the
3/2$^{-}$ states in $^{7}$Li for the clusterization $^4$He+$^3$H. One of these
functions is the wave function of the ground state (GS), the second function
is the Pauli resonance state (PR) with the energy $E$=25.810 MeV, and the third
function is the wave function of the nonresonant elastic $^4$He+$^3$H scattering
state (SC) ($E$=10.1 MeV). The main difference between the the wave functions of 
Pauli resonant and nonresonant states is the contribution of the oscillator
function with $n=0$. This function gives the largest
contribution to the wave function of the Pauli resonance states, and it has the
smallest contribution to the wave functions of the ground and continuous-spectrum states.%

%BeginExpansion
%EndExpansion

\begin{figure}[ht]
\begin{center}
\includegraphics[width=\textwidth]{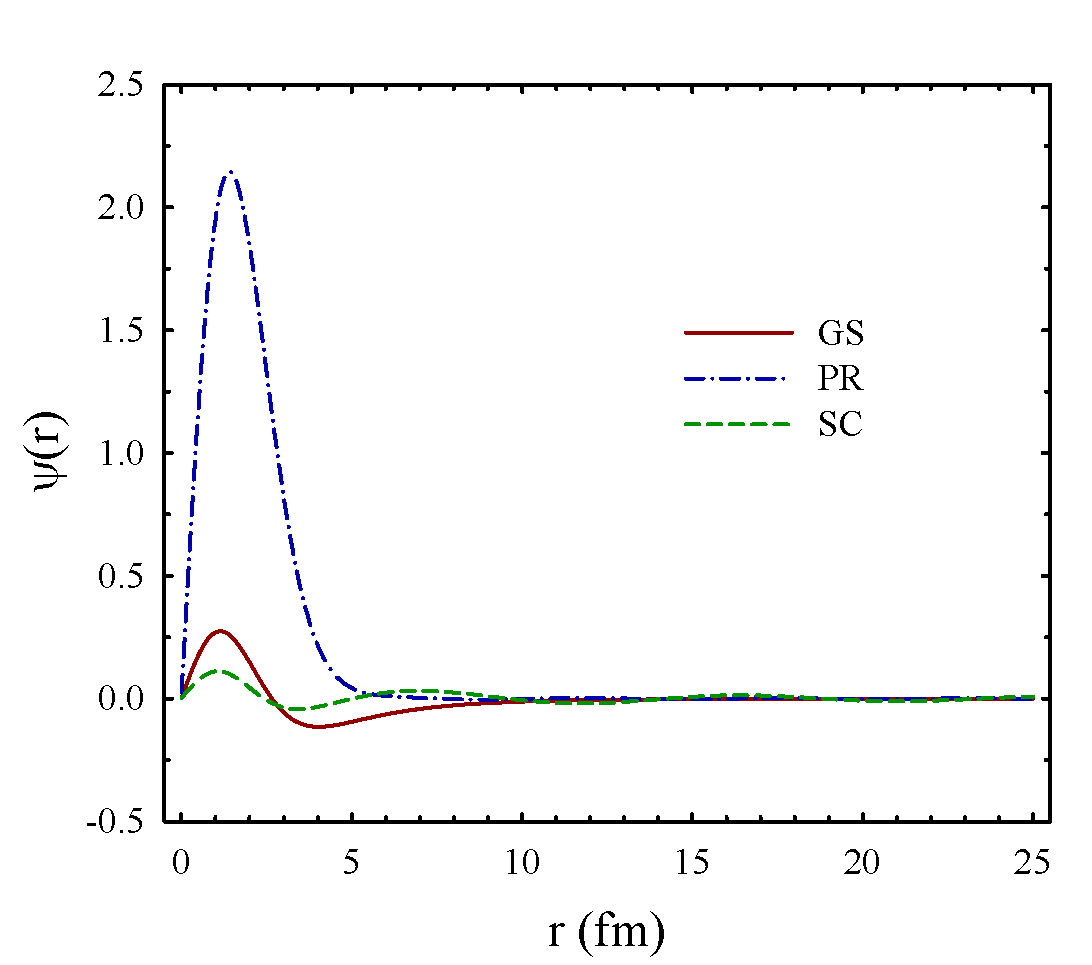}%
%{WaveFunsC32M7LiT.jpg}%
\caption{Wave functions of $^{7}$Li=$^4$He+$^3$H in the coordinate space of the
3/2$^{-}$ bound\ (BS), Pauli resonance (PR), and scattering (SC) states. 
Calculations are performed with 
the three-cluster configuration $^4$He+$d+n$.}%
\label{Fig:WaveFunsC32M7LiT}%
\end{center}
\end{figure}

In Fig. \ref{Fig:WaveFunsC32M7LiT}, we demonstrate the wave functions of these
states in the coordinate space. As one should expect, nonresonant wave functions
have a node at small distances ($r<$2.5 fm), while the resonance wave function 
has the first node at a relatively large
distance ($r\approx$5.5fm). Besides, the resonance function has a very large
amplitude, at least two times larger than the amplitude of the bound and
scattering states.

\begin{figure}[ht]
\begin{center}
\includegraphics[width=\textwidth]{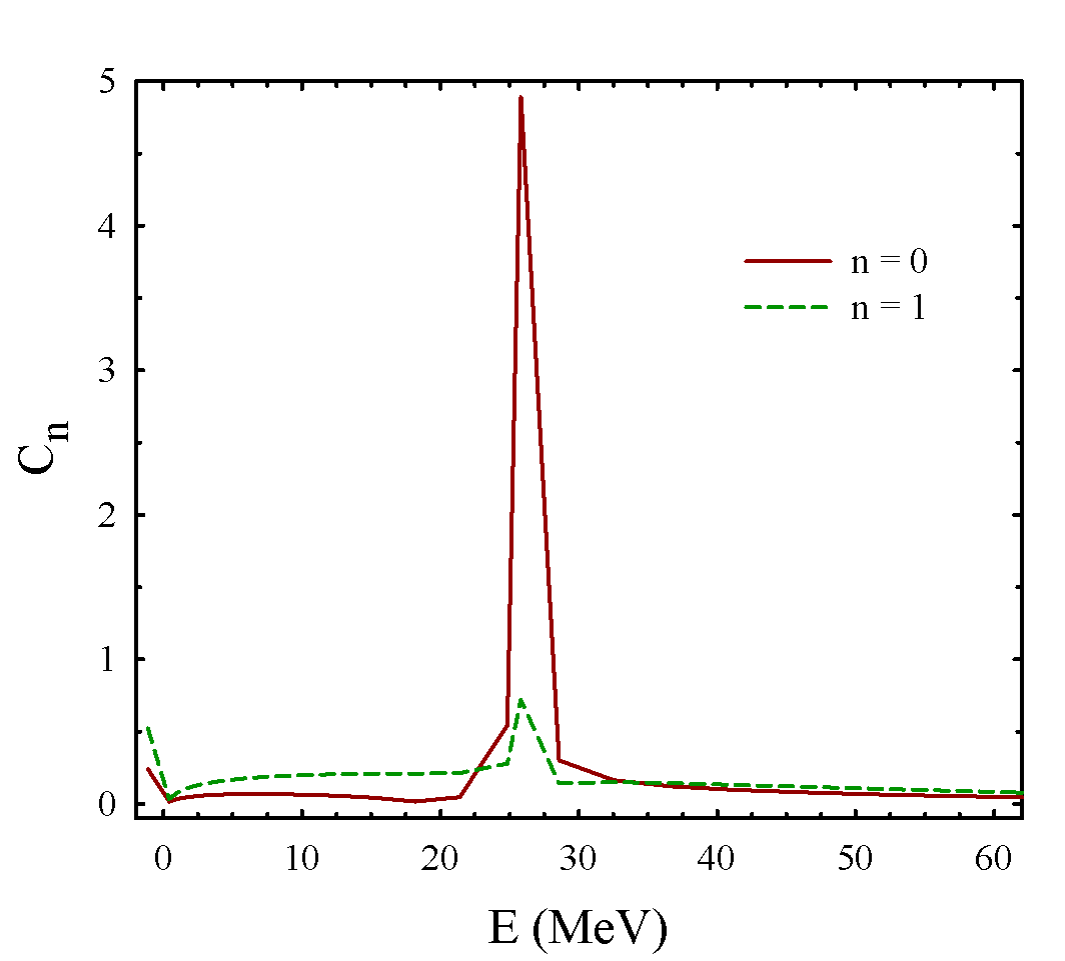}%
%{C07LiT32M.jpg}%
\caption{Contribution of the oscillator wave functions with $n=0$ and $n=1$ to the
wave functions of the continuous-spectrum 3/2$^{-}$ states of the $^4$He+$^3$H
channel.}%
\label{Fig:C07LiT32M}%
\end{center}
\end{figure}

Figure \ref{Fig:C07LiT32M} shows the general picture of the contribution of oscillator
functions with the quantum numbers $n=0$ and $n=1$ to the wave functions of
continuous-spectrum states over a large energy region. Figure \ref{Fig:C07LiT32M}
confirms also that the oscillator wave function with $n=0$ contribute mainly
to the Pauli resonance state and gives a small contribution to other states of
the $^4$He+$^3$H continuous spectrum.%

%BeginExpansion
%EndExpansion
%

%BeginExpansion
%EndExpansion

\begin{figure}[ht]
\begin{center}
\includegraphics[width=\textwidth]{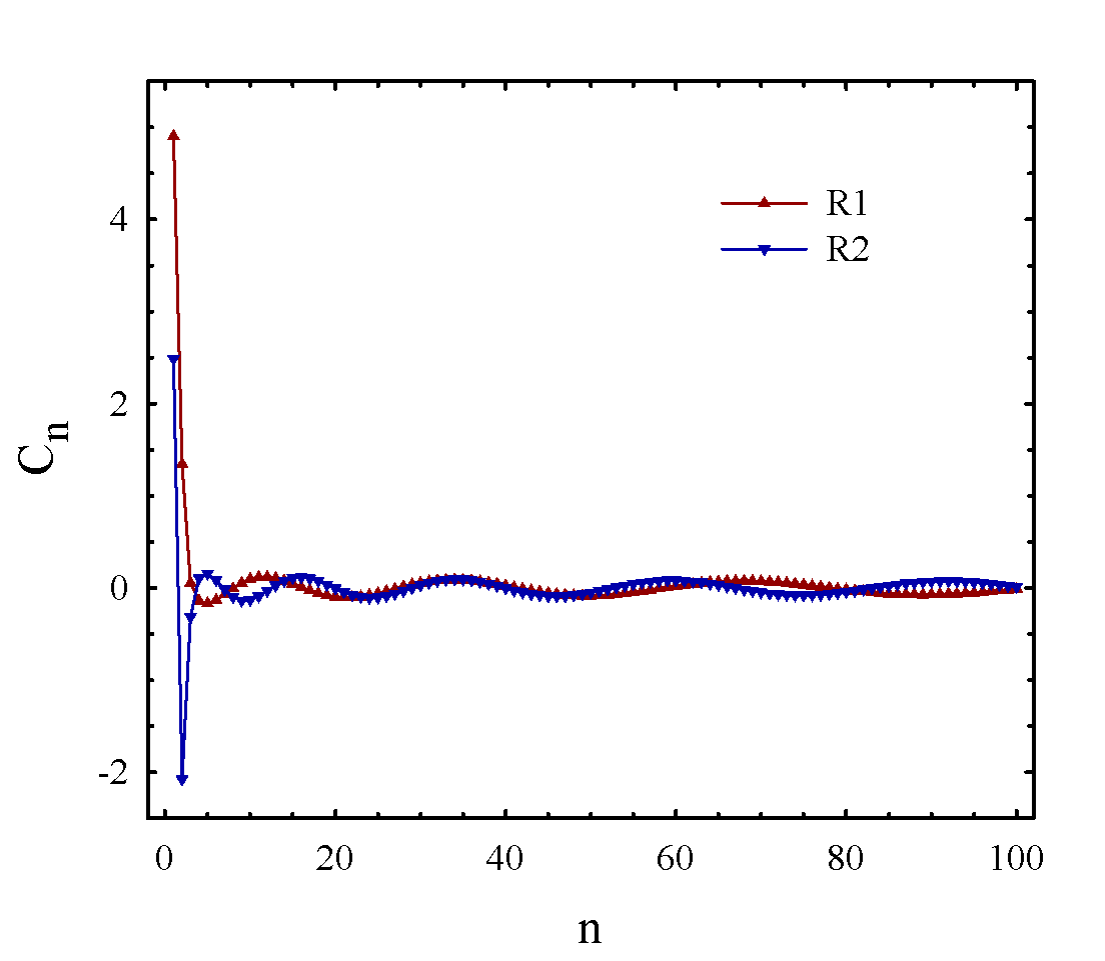}%
%{WaveFunsO12P7LiT.jpg}%
\caption{Wave functions of two Pauli resonance state in the oscillator space in
the 1/2$^{+}$ state of $^{7}$Li. Calculations are performed with 
the three-cluster configuration $^4$He+$d+n$.}%
\label{Fig:WaveFunsO12P7LiT}%
\end{center}
\end{figure}

\begin{figure}[ht]
\begin{center}
\includegraphics[width=\textwidth]{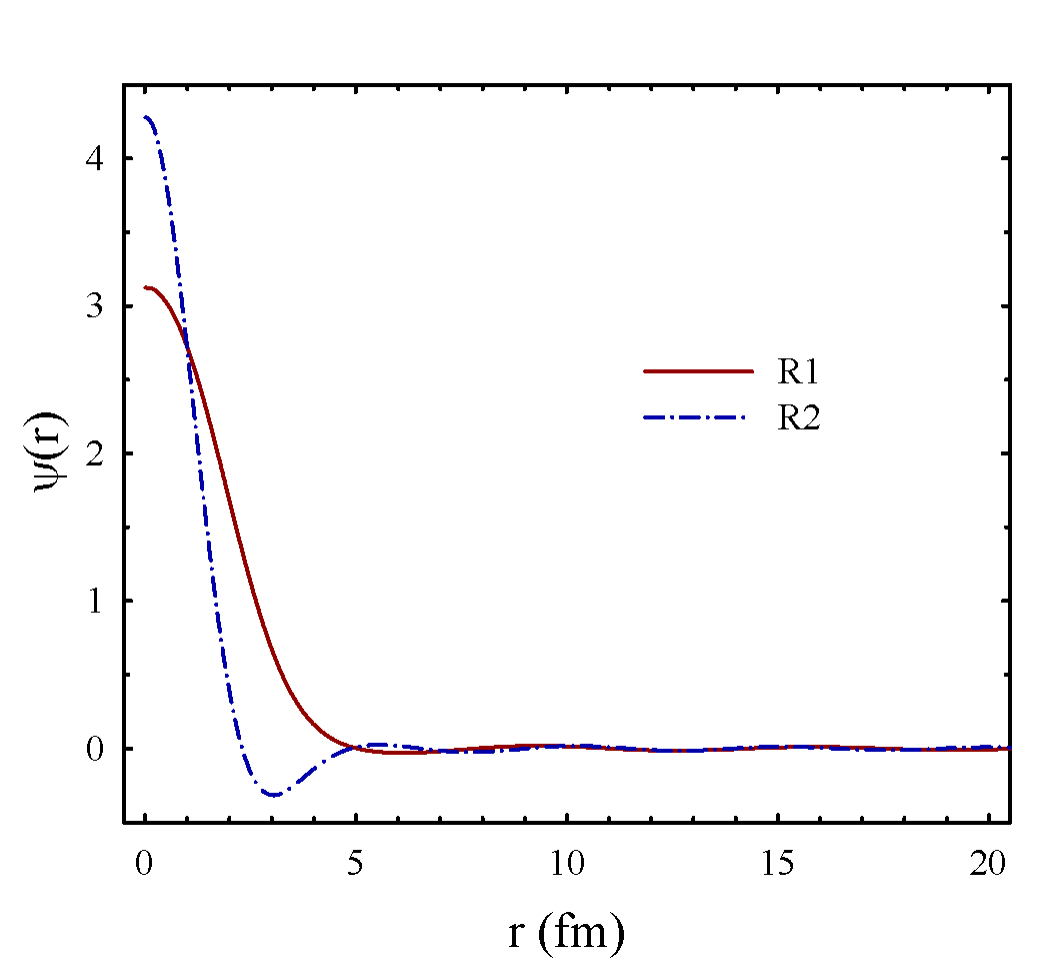}%
%{WaveFunsC12P7LiT.jpg}%
\caption{Wave functions of two Pauli resonances in the  coordinate space in the
1/2$^{+}$ state in the $^4$He+$^3$H channel. Results are obtained with the $^4$He+$d+n$ calculations.}%
\label{Fig:WaveFunsC12P7LiT}%
\end{center}
\end{figure}

Let us now consider a case with two Pauli resonance states. They are detected
in the 1/2$^{+}$ state of $^{7}$Li in the channel $^4$He+$^3$H. The wave functions
of two Pauli resonance states are shown in Fig. \ref{Fig:WaveFunsO12P7LiT} in the
oscillator space and in Fig. \ref{Fig:WaveFunsC12P7LiT} in the coordinate space.
Two oscillator functions with $n=0$ and $n=1$ give the main contribution to
the Pauli resonance functions. As a result, these wave functions in the
coordinate space describe a very compact two-cluster system, the main part of
which is concentrated at the small distances between clusters, namely, $0\leq
r\leq5$ fm. The amplitudes of the wave functions in this small region are
substantially larger than the amplitudes at large distances.%

%BeginExpansion
%EndExpansion
%

%BeginExpansion
%EndExpansion

\subsection{Overlap \label{Sec:Overlap}}

\begin{figure}[ht]
\begin{center}
\includegraphics[width=\textwidth]{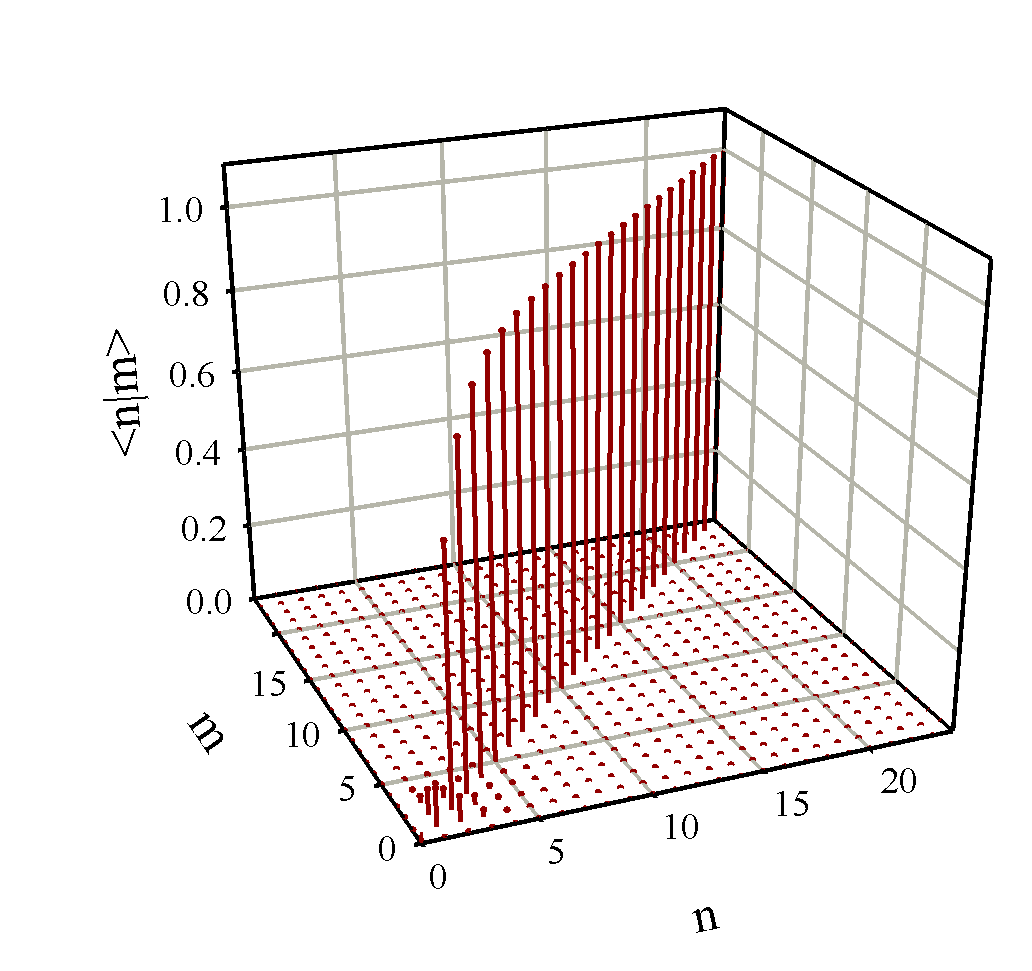}%
%{Ovlp3D7Li12P.jpg}%
\caption{Matrix for the norm kernel in the state $L$=0, $S$=1/2, and $J^{\pi}%
$=1/2$^{+}$ of the channel $^4$He+$^3$H. Results are obtained with the 
$^4$He+$d+n$ configuration.}%
\label{Fig:Ovlp3D7Li12P}%
\end{center}
\end{figure}

As it was widely recognized that the Pauli resonance states
appear due to the Pauli principle, it is then expedient to analyze its effects
on the norm kernels. Matrix of the norm kernel in general case (for the improved and
advanced versions of the RGM) is nondiagonal. Thus, we start the analysis with a
3D picture of the matrix. In Fig. \ref{Fig:Ovlp3D7Li12P}, we display the
overlap matrix $\left\Vert \left\langle n|m\right\rangle \right\Vert $ for the
channel $^4$He+$^3$H in the state $L$=0, $S$=1/2 and $J^{\pi}$=1/2$^{+}$.
\ One can see that this matrix is a quasi-diagonal. The largest matrix
elements are located on the main diagonal, and the larger is $m=n$, the closer they are
to unity. Off-diagonal matrix elements $\left\langle n|m\right\rangle
$ are very small. A few diagonal matrix elements with small values of $n$ are
also small due to the Pauli principle. One may conclude that the Pauli
principle has a short-range nature, since it affects a relatively small number of
cluster basis functions $\left\vert n\right\rangle $ determined in Eq. (\ref{eq:M11}), and the
corresponding matrix elements $\left\langle n|m\right\rangle $. Note that Fig.
\ref{Fig:Ovlp3D7Li12P} demonstrates a typical behavior of the matrix elements of a
norm kernel in the advanced version of the RGM for all nuclei and all states
considered in this paper.%

%BeginExpansion
%EndExpansion

\begin{figure}[ht]
\begin{center}
\includegraphics[width=0.75\textwidth]{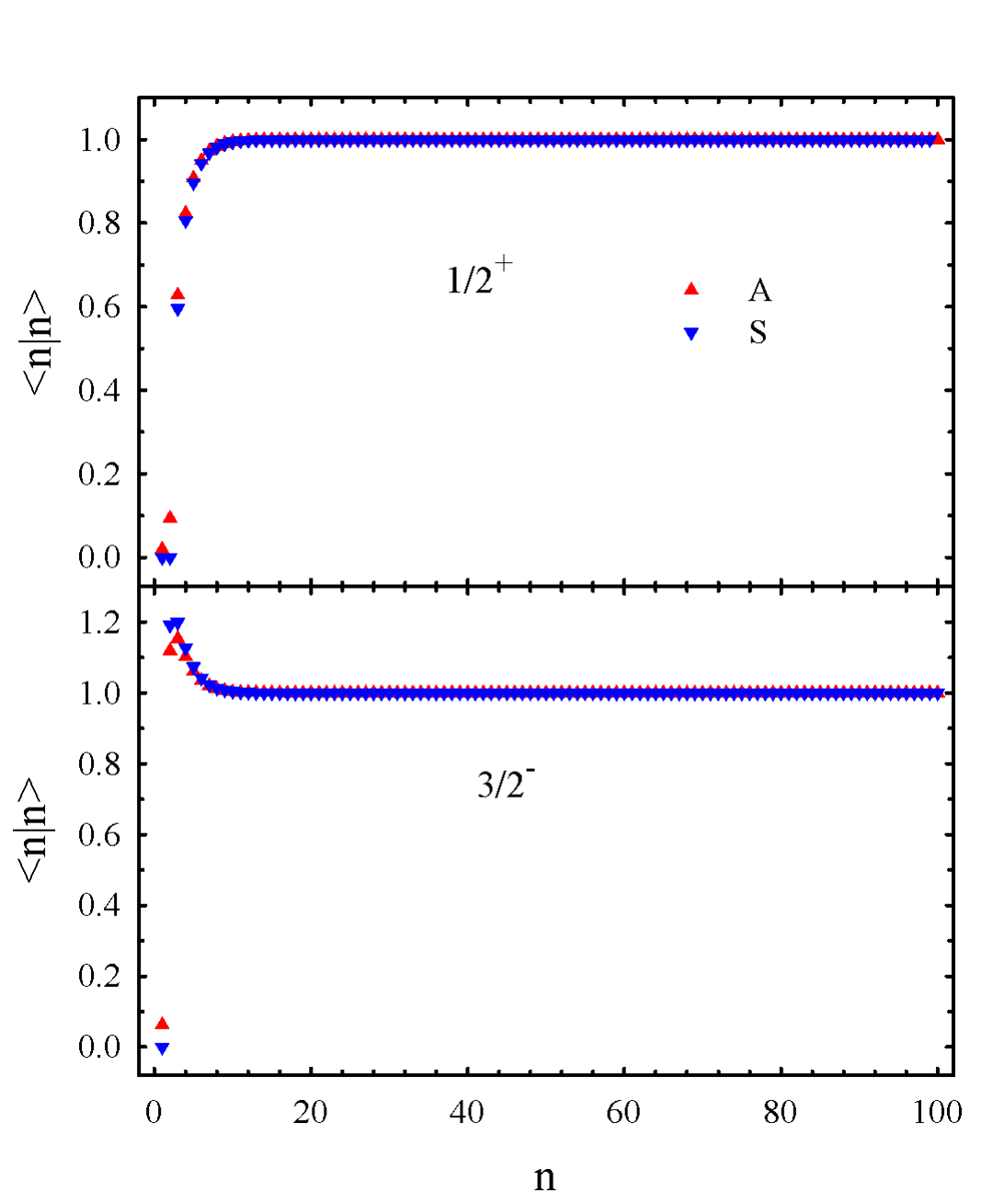}%
%{Ovlp7Li32M12P.jpg}%
\caption{Comparison of diagonal matrix elements of the norm kernel determined
in the standard (S) and advanced (A) versions of the RGM for 1/2$^{+}$ and
3/2$^{-}$ states of $^{7}$Li=$^4$He+$^3$H. Results
 are obtained with the $^4$He+$d+n$ configuration.}%
\label{Fig:Ovlp7Li32M12P}%
\end{center}
\end{figure}

Figure \ref{Fig:Ovlp3D7Li12P} prompts us to study only the diagonal matrix elements
of the norm kernel, which completely reflects  effects of the Pauli
principle. Consequently, in this subsection, we discuss the diagonal matrix
elements and eigenvalues of the norm kernel. In Fig. \ref{Fig:Ovlp7Li32M12P},
we compare the diagonal matrix elements of the norm kernel determined in the
standard (S) and advanced (A) versions of the RGM, for $^{7}$Li as a two-cluster
configuration $^4$He+$^3$H. It is worth to recall that, in the standard
version, the matrix of the norm kernel is diagonal. This figure demonstrates
general features of the quantities $\left\langle n|n\right\rangle $ and
$\Lambda_{\alpha}$ for all two-cluster systems under consideration. As was
pointed out in the previous paragraph, the major part of diagonal matrix elements
is equal to unity and only a small fraction of them differs from unity, by showing
effects of the Pauli principle. It is necessary to recall that the oscillator wave
functions with small values of the quantum number $n$ describe two clusters at the
smallest relative distance. Thus, effects of the Pauli principle for these
functions are prominent. One can see that there are two Pauli forbidden states
in the 1/2$^{+}$ state and one in the 3/2$^{-}$ state within the standard
version for $^4$He+$^3$H. In the advanced version, these basis states, namely, $\left\vert
n,L\right\rangle  = \left\vert
0,0\right\rangle $ and $\left\vert 1, 0\right\rangle $ for 1/2$^{+}$ and
$\left\vert 0, 1\right\rangle $ for 3/2$^{-}$, can be considered as
almost forbidden Pauli states, since the corresponding diagonal matrix elements
are very small ($\left\langle n|n\right\rangle <0.1$). Fig.
\ref{Fig:Ovlp7Li32M12P} demonstrates the important features of matrix elements:
the number of forbidden states in the standard version coincides with the
number of almost forbidden states in the advanced version.

%BeginExpansion
%EndExpansion

\begin{figure}[ht]
\begin{center}
\includegraphics[width=0.75\textwidth]{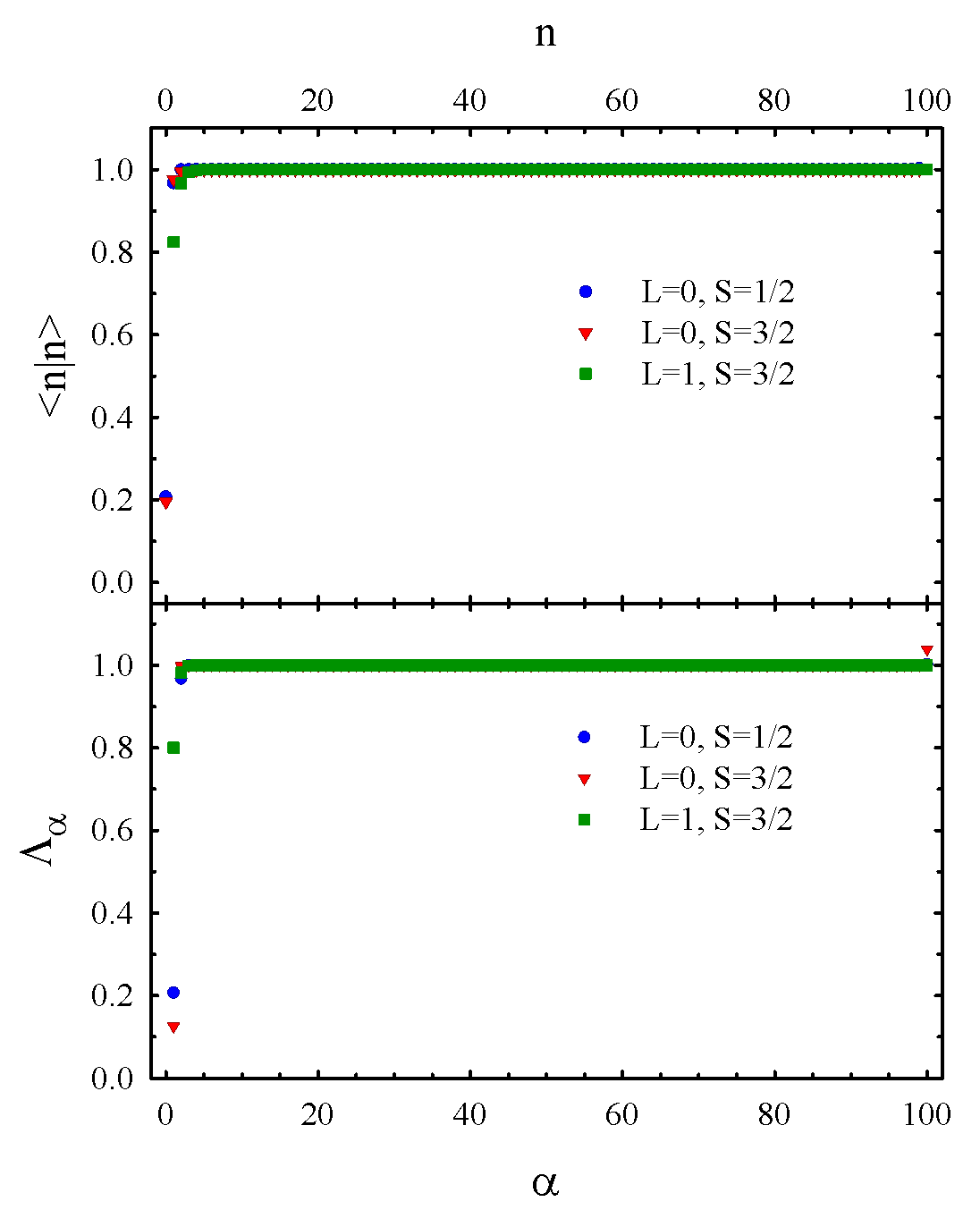}%
%{Ovlp6LiNL0L1.jpg}%
\caption{Diagonal matrix elements (the upper panel) and eigenvalues (the lower
panel) of the norm kernel for the two-cluster system $^{6}$Li+$n$. Results
 are obtained with the $^4$He+$d+n$ calculations.}%
\label{Fig:Ovlp6LiNL0L1}%
\end{center}
\end{figure}

\begin{figure}%[ht]
\begin{center}
\includegraphics[width=0.75\textwidth]{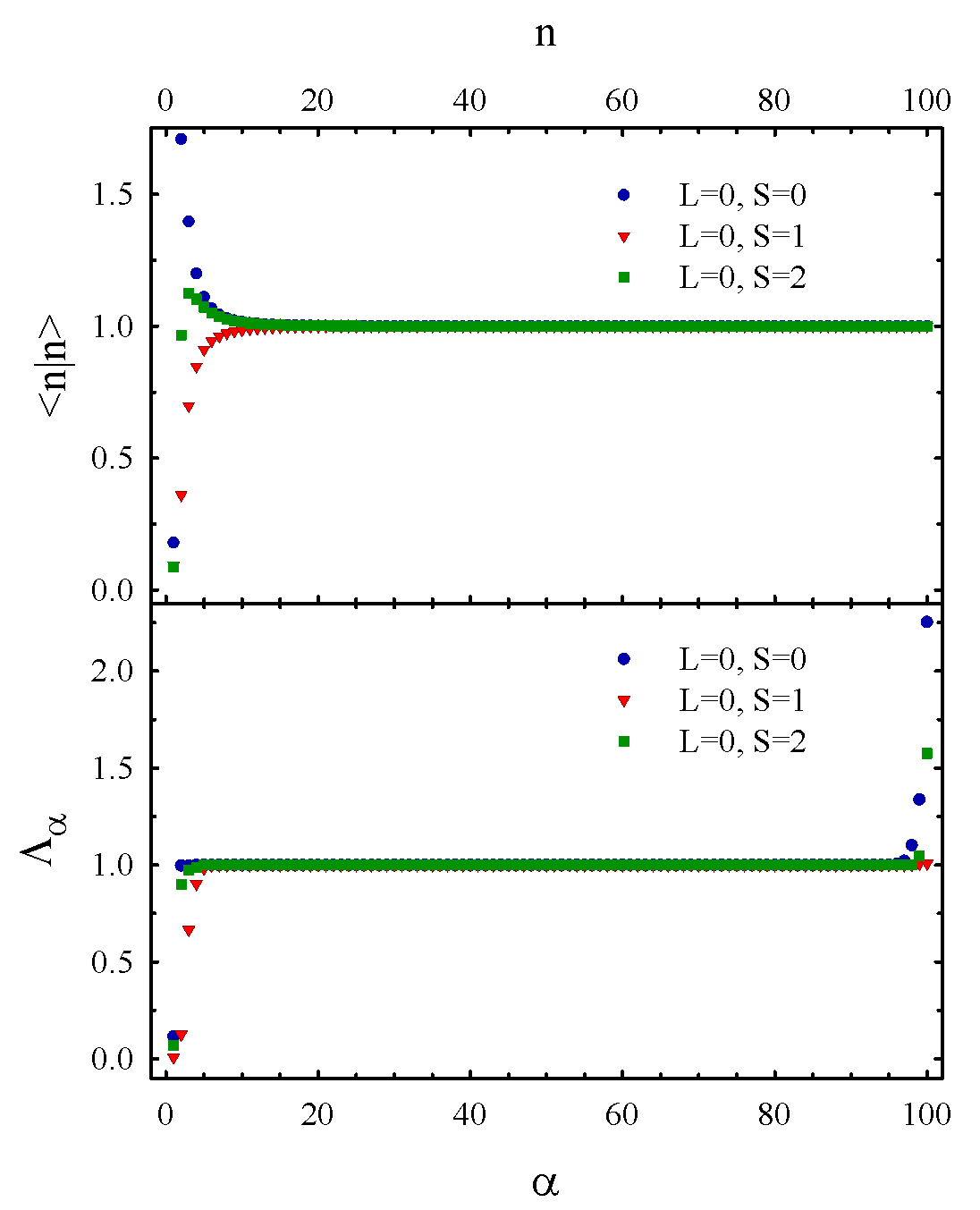}%
%{Ovlp6LiDL0.jpg}%
\caption{Diagonal matrix elements (the upper panel) and eigenvalues (the lower
panel) of the norm kernel for the $L^{\pi}$= $0^{+}$ states and different
values of the total spin $S$ in the two-cluster system $^{6}$Li+$d$. Results 
are obtained with the $^4$He+$d+d$ model space.}%
\label{Fig:Ovlp6LiDL0}%
\end{center}
\end{figure}

The diagonal matrix elements $\left\langle n|n\right\rangle $ and eigenvalues
$\Lambda_{\alpha}$ of the norm kernel for the $L^{\pi}=0^{+}$ and $1^{-}$ states 
of the advance 
two-cluster system $^{6}$Li+$n$ are shown in Fig. \ref{Fig:Ovlp6LiNL0L1}. The
almost forbidden states are found for the states $L^{\pi}=0^{+}$ and $S$=1/2
and $S$=3/2. Comparing Figs. \ref{Fig:Ovlp6LiNL0L1} and \ref{Fig:Ovlp6LiDL0}
we see that the larger are interacting clusters, the larger is the region of
diagonal matrix elements $\left\langle n|n\right\rangle $ which are affected
by the Pauli principle.

In Fig. \ref{Fig:Ovlp6LiDL0} we display the diagonal matrix elements
$\left\langle n|n\right\rangle $ and eigenvalues $\Lambda_{\alpha}$ of the
norm kernel for the $0^{+}$ states of the advanced $^{6}$Li+$d$ cluster system. Diagonal
matrix elements also show that there are a few almost forbidden states when
$\left\langle n|n\right\rangle $ are close to zero. One may observe a set of
the super-allowed Pauli states ($\left\langle n|n\right\rangle >1$) for the
total spin $S$=1 and $S$=2. There are similarities between eigenvalues and the
diagonal matrix elements of the norm kernel. The eigenvalues $\Lambda_{\alpha
}$ reveals a few almost forbidden states, two states for $S$=1 and one state
for the total spin $S$=0 and $S$=2. Similarly to the diagonal matrix elements,
the eigenvalues for $S$=0 and $S$=2 possess the super-allowed states.%

%BeginExpansion
%EndExpansion

%BeginExpansion
%EndExpansion
%

\begin{figure}[ht]
\begin{center}
\includegraphics[width=0.75\textwidth]{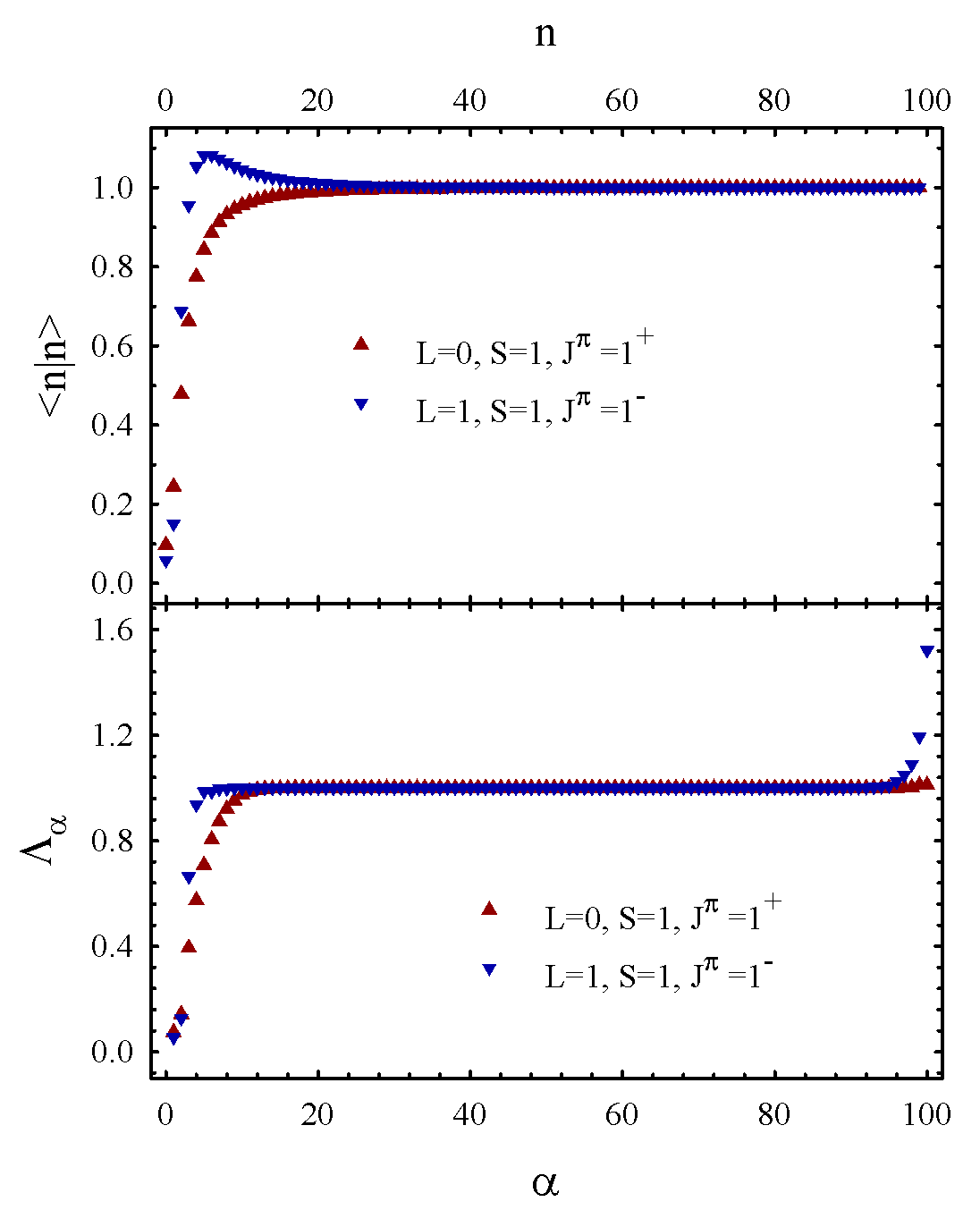}%
%{Ovlp6LiAL0L1.jpg}%
\caption{Diagonal matrix elements (upper panel) and eigenvalues (lower panel)
of the norm kernel for the channel $^{6}$Li+$^4$He in the states $L$=0,
$S$=1, $J^{\pi}$=1$^{+}$ and $L$=1, $S$=1, $J^{\pi}$=1$^{-}$. Results 
are obtained with the $^4$He+$^4$He+$d$ model space.}%
\label{Fig:Ovlp6LiAL0L1}%
\end{center}
\end{figure}

Diagonal matrix elements $\left\langle n|n\right\rangle $ of the norm kernel
and its eigenvalues $\Lambda_{\alpha}$ for the channel $^{6}$Li+$^4$He 
in the advanced two-cluster system are
displayed in Fig. \ref{Fig:Ovlp6LiAL0L1}. Two almost forbidden states are
demonstrated by both diagonal matrix elements and eigenvalues. They are
observed in two states: $L$=0, $S$=1, $J^{\pi}$=1$^{+}$ and $L$=1, $S$=1,
$J^{\pi}$=1$^{-}$.%

%BeginExpansion
%EndExpansion

Finishing this subsection, we conclude that the number of almost forbidden
states coincides with the number of almost forbidden eigenstates. Almost
forbidden states $\left\vert n\right\rangle $ obey the restriction
$\left\langle n|n\right\rangle <0.3$, while almost forbidden eigenstates have
$\Lambda_{\alpha}<0.2$. Comparing the results demonstrated in Figs.
\ref{Fig:Ovlp3D7Li12P}, \ref{Fig:Ovlp7Li32M12P}, \ref{Fig:Ovlp6LiNL0L1}, \ref{Fig:Ovlp6LiDL0}
 with the results in Tables \ref{Tab:PauliResons1} and
\ref{Tab:PauliResons2}, we came to the conclusion that the number of almost
forbidden states equals of the number of the Pauli resonance states.

\section{Method REV \label{Sec:REV}}

Let us consider the main ideas of the REV method formulated in
Ref. \cite{1992NuPhA.548...39K}. The authors of
Ref. \cite{1992NuPhA.548...39K} paid attention to a set of new eigenstates
of the norm kernel appeared in the case when different oscillator lengths were
used for an alpha particle ($b_{\alpha}$=1.395 fm) and $^{16}$O ($b_{O}$=1.776 fm). 
These eigenstates have very small
values as compared to eigenstates with the common oscillator length. For
example, the smallest eigenvalue obtained for the total orbital momentum
$L^{\pi}$=0$^{+}$ with the common oscillator length is equal to 0.229, while
there are four eigenstates with eigenvalues less than 0.03. The similar picture
was also observed for the state $L^{\pi}$=1$^{-}$. The lowest eigenvalue
obtained with the common oscillator length equals 0.344. For different
oscillator lengths, four
eigenstates emerge with eigenvalues less than 0.04.

It was suggested in Ref. \cite{1992NuPhA.548...39K} to eliminate such
eigenvalues and to use a smaller set of norm kernel eigenstates. Thus, in the
case of different oscillator lengths, all eigenstates with eigenvalues smaller
than the smallest eigenvalue with the common oscillator length were treated as the Pauli
forbidden states. Actually, the border between the Pauli allowed and Pauli
forbidden state the Pauli allowed and Pauli forbidden states in system
$^4$He+$^{16}$O was selected to be 0.1. Having applied such restrictions, all
Pauli resonance states disappeared.

We will use this method to eliminate the Pauli resonance states which
appear in light nuclei within the advanced resonating group method. The analysis
of the eigenvalues of the norm kernel carried out in Section \ref{Sec:Overlap}
indicates that we have to redetermine the border between the Pauli allowed and
Pauli forbidden states.

The efficiency of the REV method will be demonstrated in Section
\ref{Sec:Demonstration}.

\section{Method ROF \label{Sec:ROF}}

We suggest another method to struggle with the Pauli resonance states in light nuclei
 in the improved and advanced RGM calculations. This is because the Pauli 
resonances don't appear in actual nuclei, but appear in 
theoretical calculations. The method is based on properties of the matrix 
elements of the norm kernel.
By analyzing the properties of matrix elements, our attention was focused on
by behavior of diagonal matrix elements $\left\langle n|n\right\rangle $. In
many cases, the matrix element $\left\langle 0|0\right\rangle $ and, sometimes,
matrix element $\left\langle 1|1\right\rangle $ are very small with respect to
other diagonal matrix elements. The analysis also revealed that
the matrix elements of the corresponding rows ($\left\langle 0|n\right\rangle $,
$\left\langle 1|n\right\rangle $) and columns ($\left\langle n|0\right\rangle
$, $\left\langle n|1\right\rangle $) are also very small. Besides, it was
shown above (Section \ref{Sec:PropertPR}) that the oscillator functions with $n=0$
and, sometimes, with $n=1$ dominate in the wave functions of the Pauli resonance
states. Thus, we suggest to omit those parts of the matrix $\left\Vert
\left\langle n|\widetilde{n}\right\rangle \right\Vert $ whose diagonal matrix
elements are very small. We also suggest a criterion of smallness for the
diagonal matrix elements. Let us introduce the minimal value of the diagonal
matrix elements $O_{\min}$ which will mark a border between the Pauli
forbidden (or almost forbidden) and Pauli allowed states. Within our method,
all diagonal matrix elements which are smaller than $O_{\min}$ will be omitted
with their correspondent rows and columns.

The analysis of the diagonal matrix elements of the norm kernel leads us to the
conclusion that in many improved and advanced two-cluster cases, considered above, 
$O_{\min}$ can be
set to 0.2. This can be seen in Figs. \ref{Fig:Ovlp6LiDL0},
 \ref{Fig:Ovlp6LiAL0L1}. Such a value can be also used
both for the case of one or two Pauli resonance states.

It is important to notice, that from mathematical point of view almost
forbidden basis states or eigenstates are allowed states and should not create
any problems. The same is true also from computational point of view, as the
smallest eigenvalues are much larger than the smallest numerical value
(numerical zero) in modern computers. Indeed, almost forbidden states do not
create any problem for bound states and their parameters, such as
root-mean-square mass and proton radii and so on. Presence of almost forbidden
states affects (distorts) only continuous-spectrum states. In this respect,
the REF and ROF methods suggest the re-determination of essentially allowed 
Pauli states. The REF and ROF
methods determine border between almost forbidden and allowed states. This
border is marked by $\Lambda_{\min}$ and $O_{\min}$ in the REF and ROF
\ methods, respectively. In the general case, one can use $\Lambda_{\min}$ and
$O_{\min}$ as variational parameters to control the number of eliminated
basis states $\left\vert n\right\rangle $ or eigenstates $\left\vert
\alpha\right\rangle $ and their effects on scattering parameters. Naturally,
the main aim of such procedure is to eliminate the Pauli resonance state(s)
and to cause minimal effects on bound states and shape resonance states.

\subsection{Demonstration of the REV and ROF methods
\label{Sec:Demonstration}}

Having analyzed the diagonal matrix elements and eigenvalues of the overlap
matrix, we deduced $O_{\min}$ and $\Lambda_{\min}$\ for all improved and 
advanced RGM calculations and, for
those states $J^{\pi}$, which have the Pauli resonance states. These quantities
are displayed in Table \ref{Tab:OminLmin}. In this table, we also indicated 
the number $N_{f.s.}$ of eliminated basis functions or eigenfunctions.%

%TCIMACRO{\TeXButton{B}{\begin{table}[tbp] \centering}}%
%BeginExpansion
\begin{table}[ht] \centering
%EndExpansion
\begin{ruledtabular}  
\caption{Actual values of $O_{\min}$ and $\Lambda_{\min}$, and 
the number of forbidden states $N_{f.s.}$, which are omitted to 
eliminate the Pauli resonance states. \label{Tab:OminLmin}}%

\begin{tabular}
[c]{cccccccc}%\hline
Nucleus & Clusterization & $L$ & $S$ & $J^{\pi}$ & $O_{\min}$ & $\Lambda
_{\min}$ & $N_{f.s.}$\\\hline
$^{6}$Li & $^{4}$He+$d$ & 0 & 1 & $1^{+}$ & 0.2 & 0.2 & 1\\
&  & 1 & 1 & $2^{-}$ & 0.2 & 0.2 & 1\\%\cline{2-8}
& $^{3}$H+$^{3}$He & 0 & 1 & $1^{+}$ & 0.1 & 0.1 & 1\\
&  & 1 & 1 & $2^{-}$ & 0.1 & 0.1 & 1\\\hline
$^{7}$Li & $^{4}$He+$^{3}$H & 1 & 1/2 & $3/2^{-}$ & 0.1 & 0.1 & 1\\
&  & 1 & 1/2 & $1/2^{-}$ & 0.1 & 0.1 & 1\\
&  & 0 & 1/2 & $1/2^{+}$ & 0.1 & 0.1 & 2\\
&  & 2 & 1/2 & $3/2^{+}$ & 0.1 & 0.1 & 1\\
& $^{6}$Li+$n$ & 0 & 1/2 & $1/2^{+}$ & 0.3 & 0.3 & 1\\\hline
$^{8}$Be & $^{6}$Li+$d$ & 0 & 0 & $0^{+}$ & 0.2 & 0.2 & 1\\
&  & 0 & 1 & $1^{+}$ & 0.1 & 0.1 & 1\\
&  & 1 & 0 & $1^{-}$ & 0.2 & 0.2 & 1\\
&  & 1 & 1 & $2^{-}$ & 0.3 & 0.3 & 1\\
&  & 0 & 2 & $2^{+}$ & 0.1 & 0.1 & 1\\\hline
$^{9}$Be & $^{6}$Li+$^3$H & 0 & 1/2 & $1/2^{+}$ & 0.2 & 0.1 & 2\\
&  & 1 & 1/2 & $1/2^{-}$ & 0.1 & 0.1 & 1\\
&  & 0 & 3/2 & $3/2^{+}$ & 0.2 & 0.1 & 2\\
&  & 1 & 3/2 & $5/2^{-}$ & 0.1 & 0.1 & 1\\
&  & 1 & 3/2 & $5/2^{-}$ & 0.2 & 0.2 & 1\\\hline
$^{10}$B & $^{6}$Li+$^{4}$He & 1 & 1 & $0^{-}$ & 0.3 & 0.2 & 2\\
&  & 1 & 1 & $1^{-}$ & 0.3 & 0.2 & 2\\
&  & 2 & 1 & $1^{+}$ & 0.2 & 0.2 & 2\\
&  & 2 & 1 & $2^{+}$ & 0.2 & 0.2 & 1\\
&  & 2 & 1 & $3^{+}$ & 0.2 & 0.2 & 1\\%\hline
\end{tabular}
\end{ruledtabular}  
%\label{Tab:OminLmin}%
%TCIMACRO{\TeXButton{E}{\end{table}}}%
%BeginExpansion
\end{table}%
%EndExpansion

\begin{figure}[ht]
\begin{center}
\includegraphics[width=0.75\textwidth]{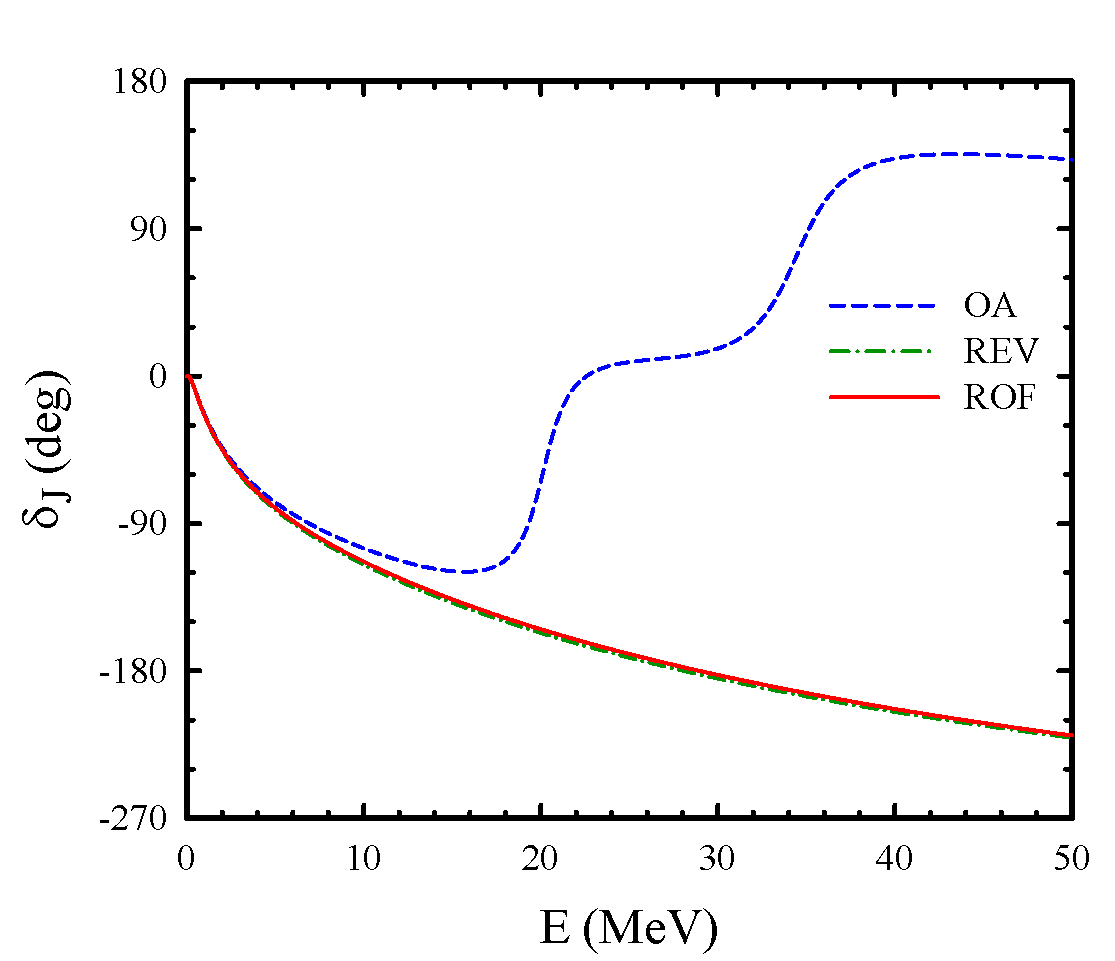}%
%{PhasesAT12P3App.jpg}%
\caption{Phase shifts of the elastic $^4$He+$^3$H scattering in the 1/2$^{+}$ state
as a function of the energy determined in three different approaches. The three-cluster 
configuration $^{4}$He+$d+n$ is involved in calculations.}%
\label{Fig:PhasesAT12P3App}%
\end{center}
\end{figure}

In Fig. \ref{Fig:PhasesAT12P3App}, we demonstrate efficiency of the REV and
ROF methods for the $^4$He+$^3$H scattering in the 1/2$^{+}$ state. Here, OA
stands for ordinary algorithm of obtaining phase shifts within the advanced
version of the RGM. The phase shift in this approach exhibits two Pauli
resonance states, parameter of which are shown in Table \ref{Tab:PauliResons1} 
and \ref{Tab:PauliResons2}. As we can see, both methods remove the Pauli 
resonance states. They also
yield the phase shifts which are close to the standard version at low energy
region 0$\leq E<$6 MeV. There is very small difference of phase shifts
obtained with the REV and ROF methods. We used minimal values of
$\Lambda_{\min}=O_{\min}=0.2$. This restriction eliminated two functions in
both methods.

%BeginExpansion
%EndExpansion

\begin{figure}[ht]
\begin{center}
\includegraphics[width=\textwidth]{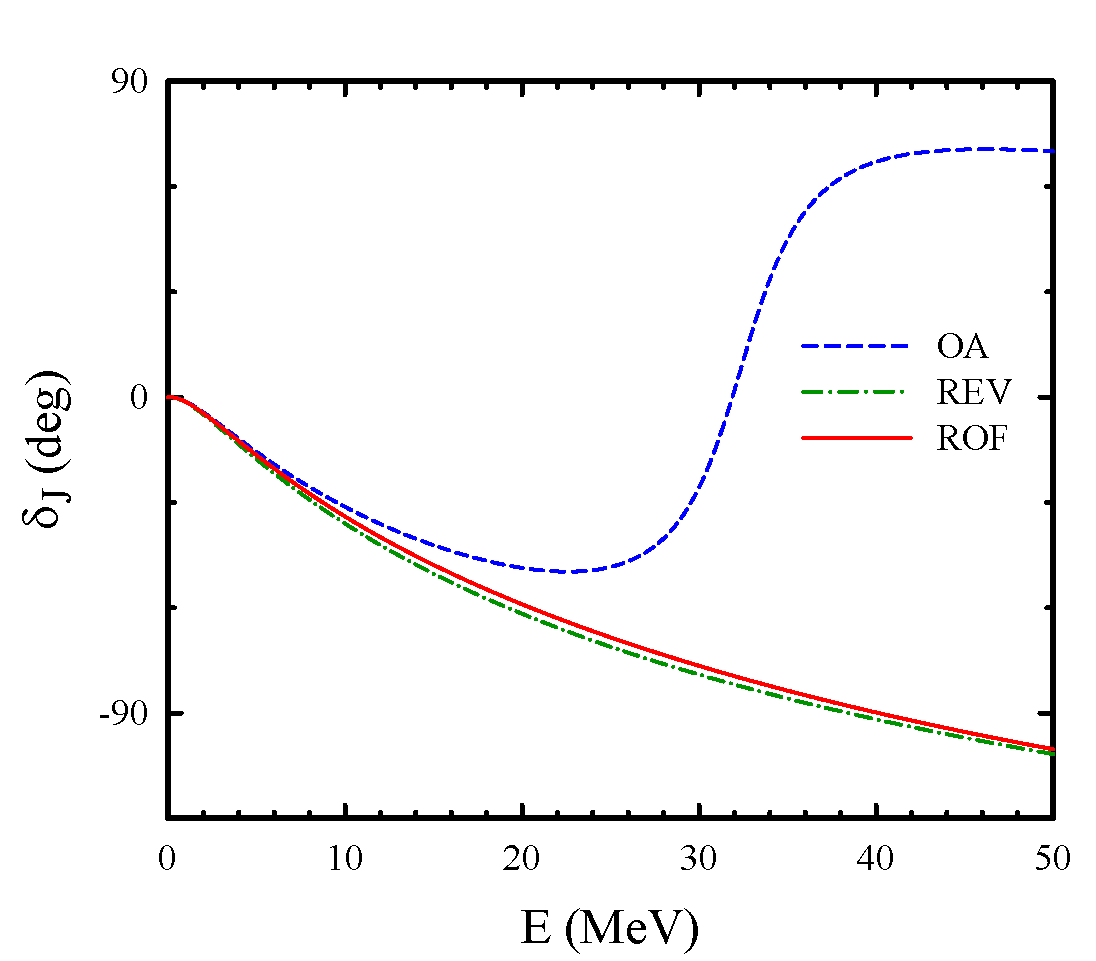}%
%{PhasesAD2M3App.jpg}%
\caption{Phase shifts of the elastic $^4$He+$d$ scattering in the 2$^{-}$
state determined in three different approaches with the three-cluster 
configuration $^4$He+$p+n$.}%
\label{Fig:PhasesAD2M3AppN}%
\end{center}
\end{figure}

The similar picture is observed for the $^4$He+$d$ scattering in the 2$^{-}$
state, see Fig. \ref{Fig:PhasesAD2M3AppN}. Only one Pauli resonance state is
generated in this case. Both REV and ROF methods remove that Pauli resonance
state and produce the phase shifts with very small differences. In this case,
we also used minimal values of $\Lambda_{\min}=O_{\min}=0.2$. This restriction
eliminated only one function in both methods.%

%BeginExpansion
%EndExpansion

\begin{figure}[ht]
\begin{center}
\includegraphics[width=\textwidth]{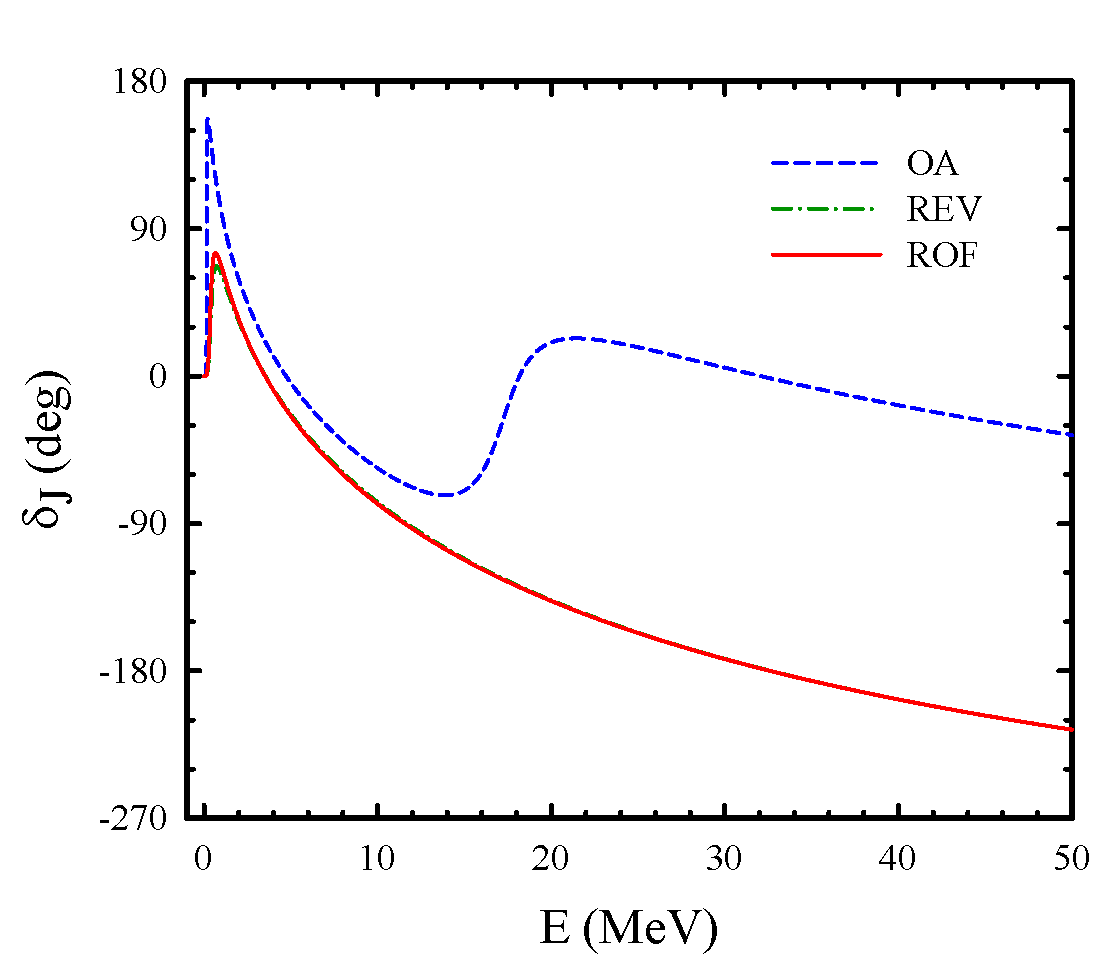}%
%{Phases6LiD0P3App.jpg}%
\caption{Phase shifts of the elastic $^{6}$Li+$d$ scattering in the state
$L=0$, $S=0$ and $J^{\pi}=0^{+}$. Calculations are performed with the 
three-cluster configuration $^{4}$He+$d$+$d$.}%
\label{Fig:Phases6LiD0P3App}%
\end{center}
\end{figure}

Phase shifts of the elastic $^{6}$Li+$d$ scattering obtained within three
different approaches are shown in Fig. \ref{Fig:Phases6LiD0P3App}. As one can
see, in this case, we observe both low-energy shape and high-energy Pauli
resonance states. The REV and ROF methods eliminating one eigenfunction and
one oscillator function, respectively, remove the Pauli resonance state. They
also slightly change parameters of the shape resonance. In the original
approach (OA), parameters of the shape resonance are $E$=0.153 MeV and
 $\Gamma$=0.013 MeV, while in the REV method they are $E$=0.374 MeV and
 $\Gamma$=0.485 MeV and, in the ROF method, we obtained $E$=0.352 MeV and
 $\Gamma$=0.371 MeV. Note that the REV and ROF give almost identical phase
shifts for the $^{6}$Li+$d$ scattering. This means that the eliminated
eigenfunction of the norm kernel and the eliminated oscillator function are
close to each other.%

%BeginExpansion
%EndExpansion

\begin{figure}[ht]
\begin{center}
\includegraphics[width=\textwidth]{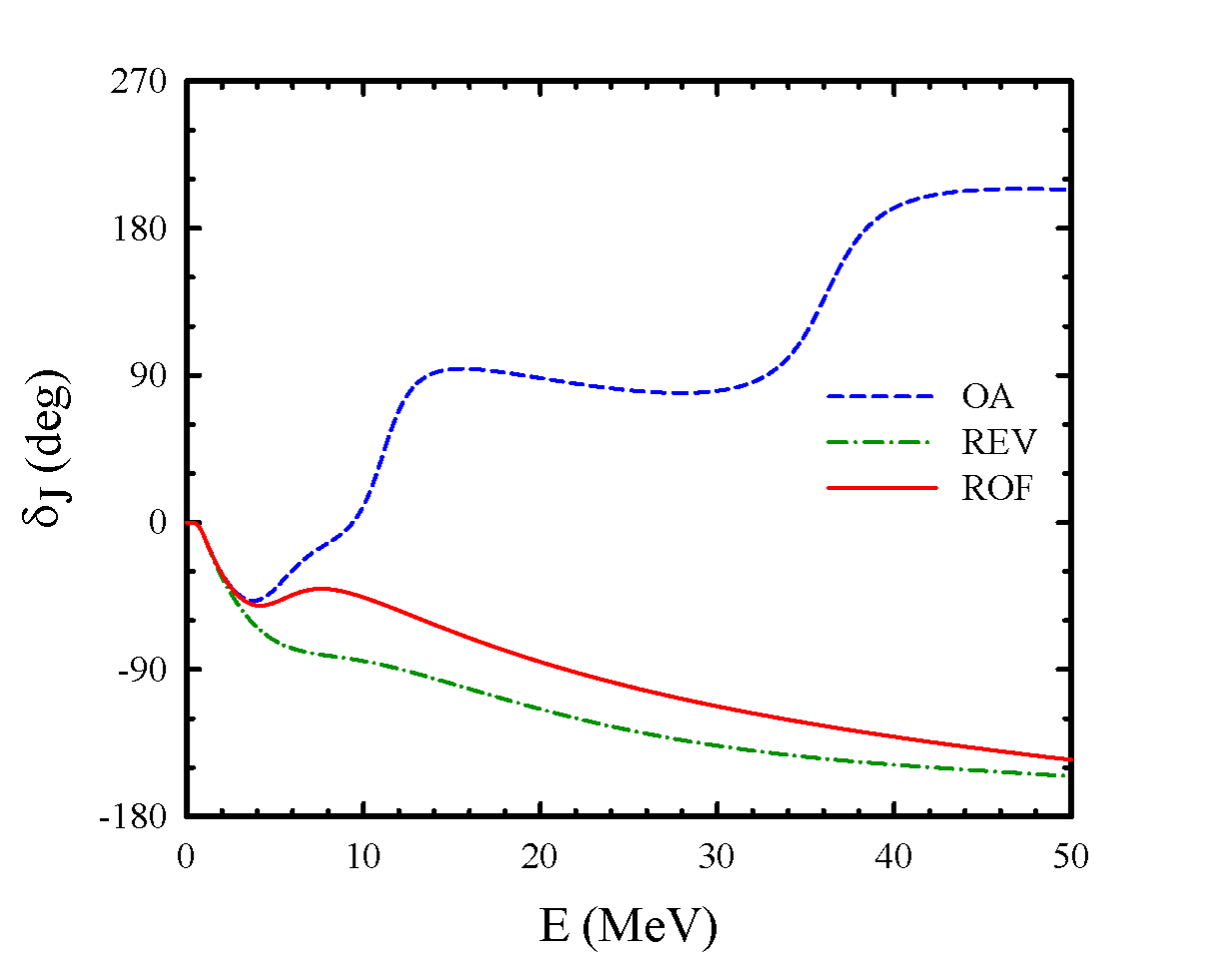}%
%{Phases6LiA1M3App.jpg}%
\caption{Phase shifts of the $^{6}$Li+$^4$He scattering in the state $L=1$,
$S=1$, $J^{\pi}=1^{-}$ obtained with three different approaches. The 
three-cluster configuration $^{4}$He+$^4$He+$d$ is used in calculations.}%
\label{Fig:Phases6LiA1M3App}%
\end{center}
\end{figure}

\begin{figure}[ht]
\begin{center}
\includegraphics[width=\textwidth]{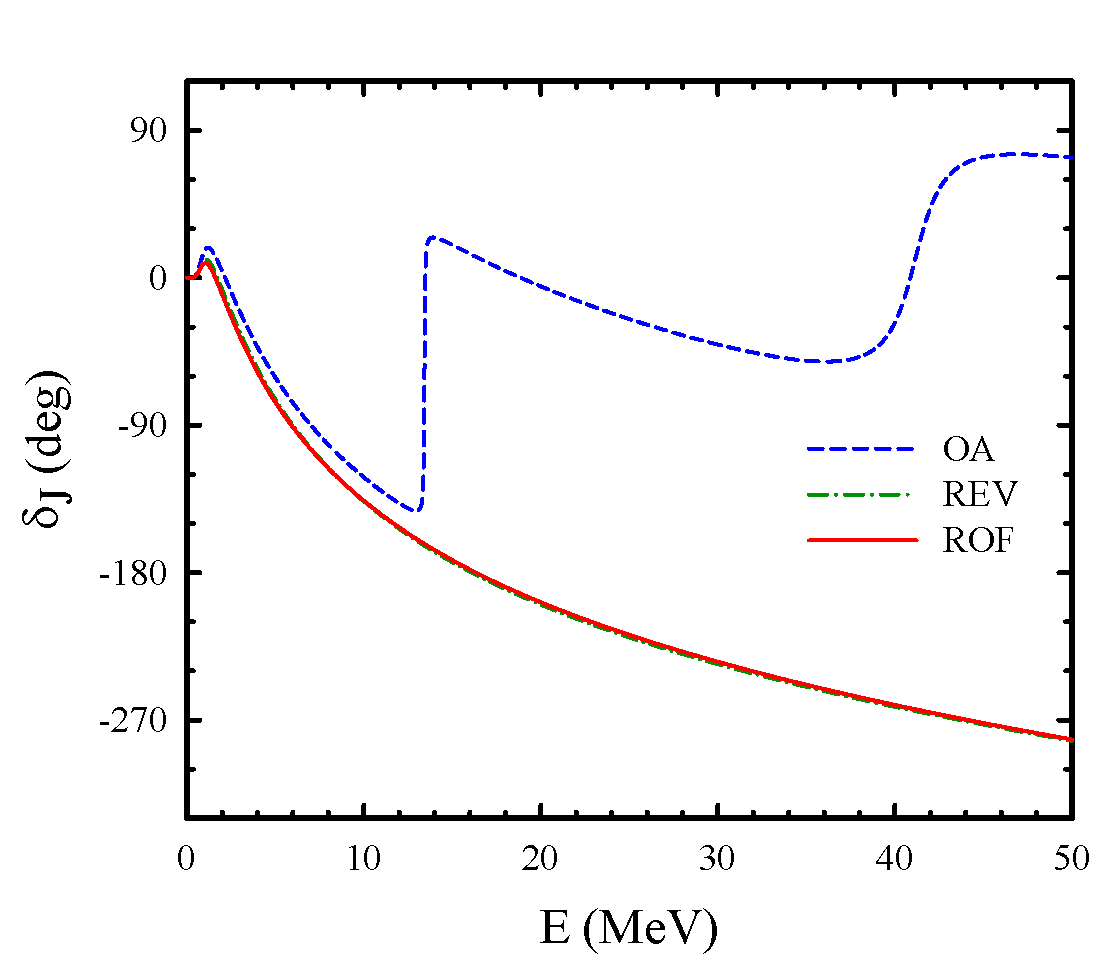}%
%{Phases6LiA1P3App.jpg}%
\caption{Phase shifts of the elastic $^{6}$Li+$^4$He scattering with
J$^{\pi}$=1$^{+}$ calculated within three approximations. Calculations 
are performed with the three-cluster configuration $^{4}$He+$^4$He+$d$.}%
\label{Fig:Phases6LiA1P3App}%
\end{center}
\end{figure}

\begin{figure}[ht]
\begin{center}
\includegraphics[width=\textwidth]{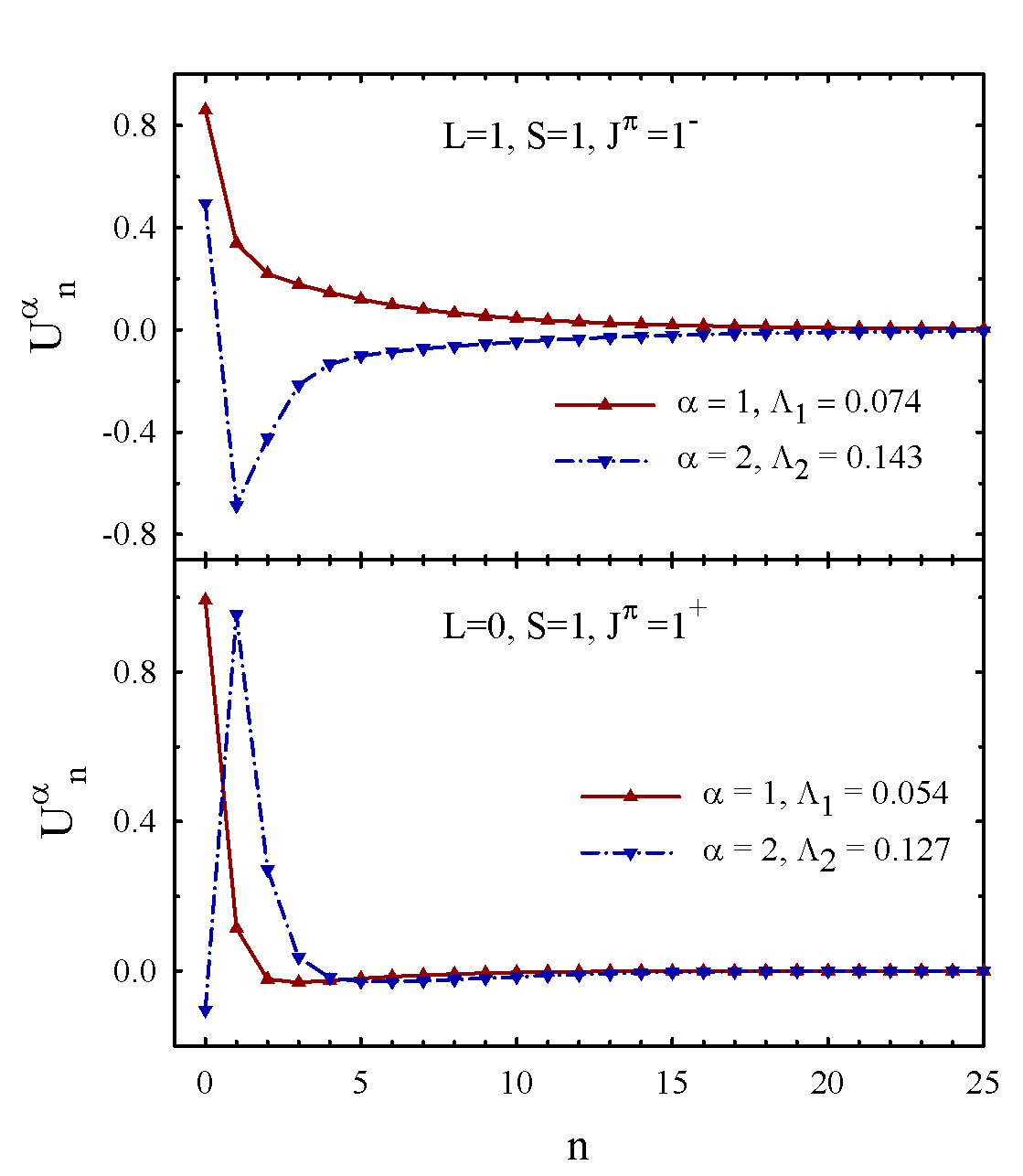}%
%{OvlpEignFuns6LiA.jpg}%
\caption{Eigenfunctions $U_{n}^{\alpha}$ of the norm kernel as a function of
$n$ for $J^{\pi}$=1$^{-}$ and $J^{\pi}$=1$^{+}$ states in $^{10}$B, considered
as a two-cluster system $^{6}$Li+$^4$He. The three-cluster configuration 
$^{4}$He+$^4$He+$d$ is involved in calculations.}%
\label{Fig:OvlpEignFuns6LiA}%
\end{center}
\end{figure}

We found several cases where the REV and ROF methods give noticeable different
phase shifts. One of such examples is shown in Fig. \ref{Fig:Phases6LiA1M3App},
where phase shifts of the $^{6}$Li+$^4$He scattering in the state $L=1$,
$S=1$ and $J^{\pi}=1^{-}$ are drawn. Note that almost the same results are
observed for the state $J^{\pi}=0^{-}$ and $J^{\pi}=2^{-}$ generated by the
coupling of the total orbital momentum $L=1$ with the total spin $S=1$. Two
Pauli resonance states were removed by eliminating two eigenfunctions of the
norm kernel obeying the restriction $\Lambda_{\alpha}\leq$0.2, and two
oscillator functions with the restriction $\left\langle n|n\right\rangle \leq
$0.3. Noticeable deviation of the phase shifts obtained in the REV and ROF
methods is seen at the energy region $E>$3 MeV.
 Such deviation can be explained by structure of the eigenfunctions and
their relation to oscillator functions. If an eigenfunction is mainly
represented by one oscillator function, one may expect close results of both
methods. If eigenfunction is spread over large number of oscillator functions,
results obtained with these two methods would be different. To prove this
statement, we show in Fig. \ref{Fig:OvlpEignFuns6LiA} eigenfunctions
$\left\Vert U_{n}^{\alpha}\right\Vert $of the norm kernel as a function of $n$
for two different cases with two Pauli resonance states. We selected cases for
elastic $^{6}$Li+$^4$He scattering with quantum numbers $L=S$=1, $J^{\pi}%
$=1$^{-}$and $L=0,$ $S$=1, $J^{\pi}$=1$^{+}$. The phase shifts for them are shown
in Figs \ref{Fig:Phases6LiA1M3App} and \ref{Fig:Phases6LiA1P3App}. Figure
\ref{Fig:OvlpEignFuns6LiA} demonstrates that, for the $J^{\pi}$=1$^{-}$ state,
a large number of oscillator functions participate in the formation of
eigenfunctions $U_{n}^{1}$ and $U_{n}^{2}$, while, for the $J^{\pi}$=1$^{+}$
state, the lowest oscillator functions with $n=0$ and $n=1$ totally dominate in the
corresponding eigenfunctions $U_{n}^{1}$ and $U_{n}^{2}$. Similar dominance of
oscillator function with the quantum number $n=0$ in the eigenfunction
$U_{n}^{1}$ are observed in all cases, when phase shifts obtained with the REV
and ROF are coincide.%

%BeginExpansion
%EndExpansion
%

%BeginExpansion
%EndExpansion
%

%BeginExpansion
%EndExpansion

\begin{figure}
[ptb]
\begin{center}
\includegraphics[width=\textwidth]{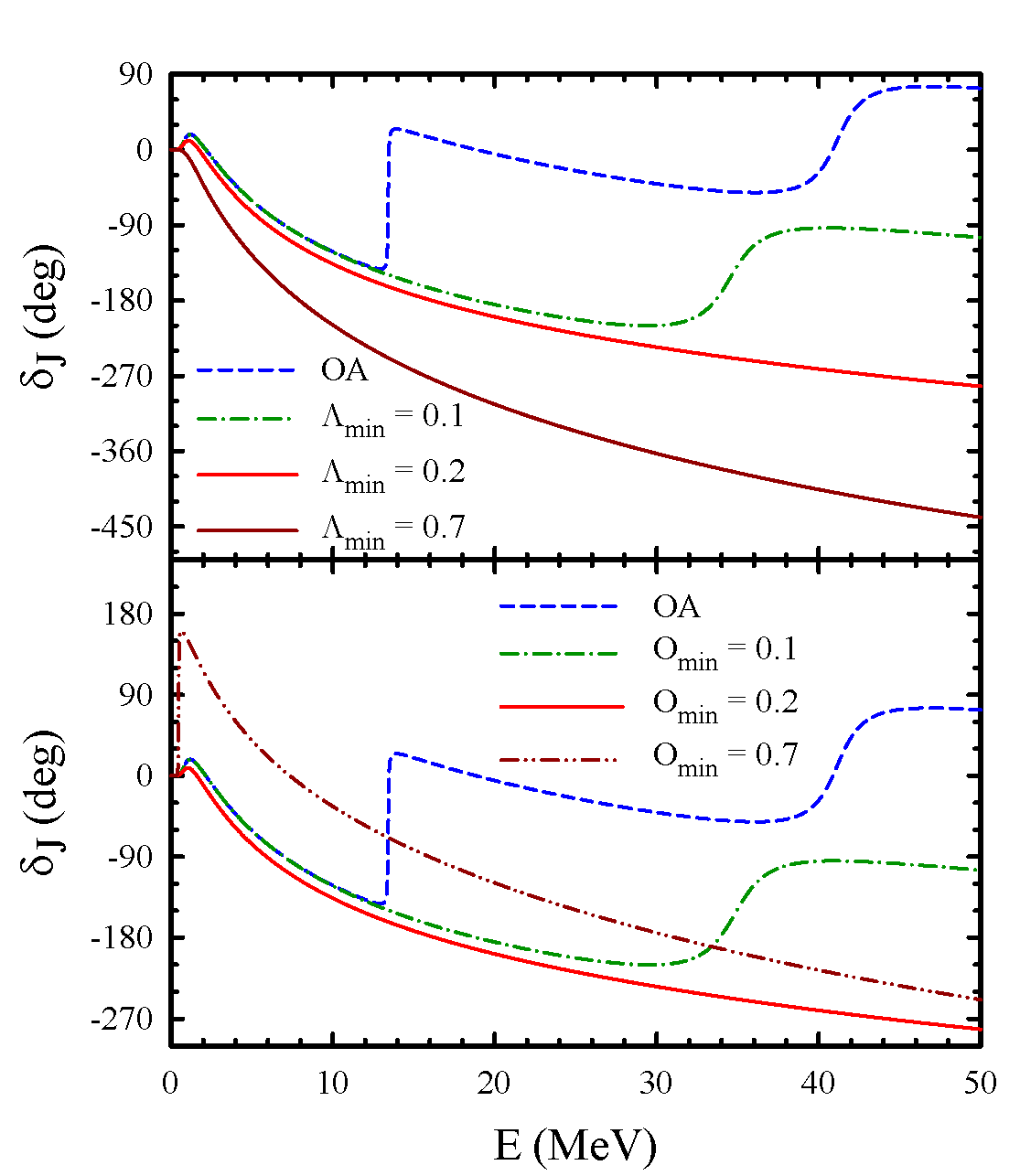}%
%{Phases6LiA1PConv.jpg}%
\caption{Phase shifts of the elastic $^{6}$Li+$^4$He scattering obtained
with different values of $\Lambda_{\min}$ (the upper part) and $O_{\min}$ (the
lower part.) Calculations are performed with the three-cluster configuration $^{4}$He+$^4$He+$d$.}%
\label{Fig:Phases6LiA1PConv}%
\end{center}
\end{figure}

In Table  \ref{Tab:Reson10B1pEvol} we show effects of eliminated
eigenfunctions and oscillator functions on parameters of bound and resonance
states. These results are obtained for the 1$^{+}$ states in $^{10}$B. By
increasing $\Lambda_{\min}$ ($O_{\min})$ from zero to a certain value,
indicated in the second column of Table \ref{Tab:Reson10B1pEvol}, we manage to
eliminate one, two and three eigenfunctions (oscillator functions). In the
fourth column of Table \ref{Tab:Reson10B1pEvol}, we demonstrate how
eliminated eigenfunctions and oscillator functions affect the energy of the
$1^{+}$ bound state of $^{10}$B. In Fig. \ref{Fig:Phases6LiA1PConv} we show
effects of eliminated functions on the $^{6}$Li+$^4$He phase shift. By
eliminating one eigenfunction or one oscillator function, we remove the lowest
Pauli resonance state and change position (lower down) of the second resonance
on approximately 6.5 MeV. However, the energy of the ground state is slightly
changed after removing one function. When we remove two eigenfunctions or two
oscillator functions, both Pauli resonance states are disappeared. Two removed
eigenfunctions increase  the energy of the bound state by 0.9 MeV, while two
removed oscillator functions increase the energy by $\approx$1.3 MeV.%

%TCIMACRO{\TeXButton{B}{\begin{table}[tbp] \centering}}%
%BeginExpansion
\begin{table}[ht] \centering
%EndExpansion
\begin{ruledtabular}  
\caption{Evolution of the $1^+$ bound and resonance states in $^{10}$B. The energies 
and widths are in MeV. \label{Tab:Reson10B1pEvol}}%
\begin{tabular}
[c]{cccccccc}%\hline
Method & $\Lambda_{\min}/O_{\min}$ & $N_{f}$ & $E_{GS}$ & $E$ & $\Gamma$ & $E$
& $\Gamma$\\\hline
OA & 0.0 & 100 & -2.477 & 13.427 & 0.056 & 41.144 & 2.751\\\hline
REV & 0.1 & 99 & -2.280 & - & - & 34.539 & 3.503\\
ROF & 0.1 & 99 & -2.264 & - & - & 34.826 & 3.684\\\hline
REV & 0.2 & 98 & -1.527 & - & - & - & -\\
ROF & 0.2 & 98 & -1.183 & - & - & - & -\\\hline
REV & 0.7 & 97 & -0.132 & - & - & - & -\\
ROF & 0.7 & 97 & 0.526 & - & - & - & -\\%\hline
\end{tabular}
\end{ruledtabular}  
%\label{Tab:Reson10B1pEvol}%
%TCIMACRO{\TeXButton{E}{\end{table}}}%
%BeginExpansion
\end{table}%
%EndExpansion
%

%BeginExpansion
%EndExpansion

As we indicated above, the oscillator functions with small values of the quantum
number $n$ and eigenfunctions with small values of index $\alpha$ describe the
most compact two-cluster configurations. It is interesting to analyze effects
of their deletion on the energies of bound states and a shape resonances, if they
appear. For this aim, we collected in Table \ref{Tab:SpectrEff} the energies of
bound and resonance states. In many cases, the elimination of the oscillator functions
  leads approximately to the same results as with the elimination of the eigenfunctions.%

%TCIMACRO{\TeXButton{B}{\begin{table}[tbp] \centering}}%
%BeginExpansion
\begin{table}[ht] \centering
%EndExpansion
\begin{ruledtabular}  
\caption{Effects of removed eigenfunctions and oscillator functions 
on the energies of bound and shape resonance states. \label{Tab:SpectrEff}}%
\begin{tabular}
[c]{ccccccc}%\hline
Nucleus & Channel & $J^{\pi}$ & Parameter & OA & REV & ROF\\\hline
$^{7}$Li & $^4$He+$^3$H & 3/2$^{-}$ & $N_{f}$ & 100 & 99 & 99\\
&  &  & $E_{GS}$, MeV & -1.127 & -1.105 & -1.064\\\hline
$^{8}$Be & $^{6}$Li$+d$ & 0$^{+}$ & $N_{f}$ & 100 & 99 & 99\\
&  &  & $E_{GS}$, MeV & -18.971 & -15.281 & -14.426\\
&  &  & $E$, MeV & 0.153 & 0.374 & 0.352\\
&  &  & $\Gamma$, MeV & 0.013 & 0.485 & 0.371\\
&  & 2$^{-}$ & $N_{f}$ & 100 & 99 & 99\\
&  &  & $E$, MeV & 0.800 & 0.825 & 0.805\\
&  &  & $\Gamma$, MeV & 0.738 & 0.957 & 0.762\\\hline
$^{10}$B & $^{6}$Li+$^4$He & 1$^{+}$ & $N_{f}$ & 100 & 98 & 98\\
&  &  & $E_{GS}$, MeV & -2.477 & -1.527 & -1.183\\%\hline
\end{tabular}
\end{ruledtabular}  
%\label{Tab:SpectrEff}%
%TCIMACRO{\TeXButton{E}{\end{table}}}%
%BeginExpansion
\end{table}%
%EndExpansion

\subsubsection{Preliminary conclusions}

At the end of this section we made preliminary conclusions concerning the REV
and ROF methods. In all cases, presented above, both methods completely remove
all detected Pauli resonance states. In many cases, both methods give close
results for phase shifts. In some cases, phase shifts are somewhat 
different. Such a difference, as we demonstrated, appear, when eigenfunctions of
the norm kernel are spread over a large number of oscillator functions. In
other words, removed eigenfunctions and removed oscillator functions are quite
different. When results of both methods coincide, the removed eigenfunctions are
presented mainly by removed oscillator functions.

We demonstrated that, the ROF method formulated in this paper, is an
alternative method to the one suggested by Kruglanski and Baye. Advantage of
the ROF is that it does not require an orthogonalization procedure of matrices
of norm kernel and then a transformation of matrix of the Hamiltonian to a new
representation. This procedure is time-consuming when a large number of basis
functions are involved. We also demonstrated that oscillator representation is
appropriate tool for studying effects of the Pauli principle on kinematic
(matrix of the norm kernel) and the dynamics (matrix of the Hamiltonian) of two- and
many-cluster systems.

\section{Conclusions \label{Sec:Conclus}}

Properties of Pauli resonance states in the two-body continuum of the light nuclei
$^{6}$Li, $^{7}$Li, $^{7}$Be, $^{9}$Be and $^{10}$B have investigated within
the advanced version of the resonating group method. The advanced version
employs a three-cluster configuration which allows one to consider in general case
three two-body (binary) channels. One of constituents of a binary channel
is considered as a two-cluster subsystem which provides us with a more correct
description of the nuclei having distinct two-cluster structures and a small
separation energy. The wave functions of two-cluster subsystem are obtained by
solving appropriate Schr\"{o}dinger equation. The advanced version we have
employed make use of the square-integrable bases - Gaussian and oscillator
bases. Gaussian basis is used to describe relative motion of two clusters in
two-cluster subsystem and is very efficient in obtaining wave functions of
bound states with a minimal number of basis functions. The Oscillator basis is used
to study interaction of the third cluster with two-cluster subsystem. It
allows us to implement proper boundary condition for discrete and continuous
spectrum states. It was demonstrated that oscillator basis is suitable tool to
study effects of the Pauli principle and to reveal nature of the Pauli
resonance states.

It was demonstrated that the advanced form of a two-cluster subsystem is the
origin of the Pauli resonance states. More precisely, an advanced form of wave
function of two-cluster subsystem is responsible for appearance of the Pauli
resonance states.

It has been shown that the Pauli resonance states appear at the relatively
high energy $E>$11 MeV. %
Some of these resonance states are very narrow resonance states,
however, major part of them are broad resonance states. The most populated
area of resonance states lies in the interval 16$<E<$21 MeV. Two dense area of
widths of resonance states are located in intervals 0.008$<\Gamma<$0.22 MeV
and 0.9$<\Gamma<$1.2 MeV.

It was found that the oscillator functions with minimal value of the quantum
number $n$ (the number of radial oscillator quanta) dominates in resonance
wave functions. These basis functions yield very small values of the diagonal
matrix elements $\left\langle n|n\right\rangle $ of the norm kernel. It was
also demonstrated that the very narrow Pauli resonance states can be detected
by using a very small number of oscillator functions: from three to five functions.
%
%Role of Pauli super-allowed and almost-forbidden states has thoroughly
%studied. New method for eliminating the Pauli resonance states is formulated
%and its efficiency is demonstrated for nuclei under consideration.

We have established that the Pauli principle predetermine appearance of the
Pauli resonance states by creating almost forbidden states, however energies
and widths of the Pauli resonance states are mainly formed by
nucleon-nucleon forces.

We found that the number of Pauli resonance states for the given $J^{\pi}$
state, discovered within the advanced version of the RGM, coincides with the
number of the Pauli forbidden states determined in the standard version of the RGM.

One of the main conclusions of the present paper is that one needs to find the  
proper definition of the Pauli forbidden and Pauli allowed (fully or
partially) states. Standard or a formal definition for Pauli forbidden states is
that the eigenvalues for them should be equal zero $\Lambda_{\alpha}=0$. Then
the Pauli allowed states should have $\Lambda_{\alpha}>0$. However, the
carried out analysis leads us to the conclusion that, for light nuclei with the
two-body clusterization, the border between forbidden and allowed states is
$\Lambda_{\min}=0.2$. It was also shown that oscillator functions $\left\vert
n\right\rangle $ which generate the diagonal matrix elements of the norm
kernel $\left\langle n|n\right\rangle \leq O_{\min}=0.2$, can be considered as
the Pauli forbidden states. By removing of the Pauli forbidden states, one
eliminates the Pauli resonance states and causes minor effects on the energy of
bound states and the energy and width of the shape resonance states, if they
exist. We have not found universal values of $\Lambda_{\min}$ and $O_{\min}$
for all light nuclei, which have been considered.

As for perspective of this work. In the present paper, we have restricted
ourselves to the single-channel approximation to reveal the Pauli resonance
states and find main factors responsible for the formation of such states. In the
future, we are planning to consider the appearance of the Pauli resonance
states in many-channel systems and how the REF and ROF can help to eliminate
them. Many-channel cases are specially interesting since small eigenvalues of
the norm kernel can appear due to a   strong overlap of basis functions belonging
to different channels. This strong coupling is not directly related to the
Pauli principle. This makes the problem more attractive and challenging.

\begin{acknowledgments}

We would like to thank K. Kat\={o} for stimulating discussions and encouraging
support. This work was supported in part by the Program of Fundamental
Research of the Physics and Astronomy Department of the National Academy of
Sciences of Ukraine (Project No. 0122U000889) and by the Ministry of Education
and Science of the Republic of Kazakhstan, Research Grant IRN: AP 09259876.
V.V. is grateful to the Simons foundation for financial support.

\end{acknowledgments}

\end{document}